\documentclass{aa}  

\usepackage{graphicx}
\usepackage{txfonts}

\usepackage{placeins}
\usepackage{natbib,twoopt}
\usepackage[breaklinks=true]{hyperref} 
\bibpunct{(}{)}{;}{a}{}{,}             
\makeatletter
  \newcommandtwoopt{\citeads}[3][][]{\href{http://adsabs.harvard.edu/abs/#3}
    {\def\hyper@linkstart##1##2{}
     \let\hyper@linkend\@empty\citealp[#1][#2]{#3}}}
  \newcommandtwoopt{\citepads}[3][][]{\href{http://adsabs.harvard.edu/abs/#3}
    {\def\hyper@linkstart##1##2{}
     \let\hyper@linkend\@empty\citep[#1][#2]{#3}}}
  \newcommandtwoopt{\citetads}[3][][]{\href{http://adsabs.harvard.edu/abs/#3}
    {\def\hyper@linkstart##1##2{}
     \let\hyper@linkend\@empty\citet[#1][#2]{#3}}}
  \newcommandtwoopt{\citeyearads}[3][][]
    {\href{http://adsabs.harvard.edu/abs/#3}
    {\def\hyper@linkstart##1##2{}
     \let\hyper@linkend\@empty\citeyear[#1][#2]{#3}}}
\makeatother

\begin{document}

   \title{Interaction of the relativistic jet and the narrow-line region of \\ PMN J0948+0022} 

   \author{B. Dalla Barba \inst{1,2,3};
        		M. Berton \inst{1};
		L. Foschini \inst{3};
		E. Sani \inst{1};
		L. Crepaldi \inst{4};
		E. Congiu \inst{1};
        		G. Venturi \inst{5,6}
		W.J. Hon \inst{7};
		and A. Vietri \inst{4,8}.
		}

   	\institute{ 	$^1$ European Southern Observatory (ESO), Alonso de Córdova 3107, Vitacura Santiago, Chile; \\
			$^2$ Università degli studi dell’Insubria, Via Valleggio 11, Como 22100, Italy; \\
			$^3$ Osservatorio Astronomico di Brera, Istituto Nazionale di Astrofisica (INAF), Via E. Bianchi 46, Merate (LC) 23807, Italy; \\
        			$^4$ Università degli studi di Padova, Vicolo dell'Osservatorio 3, Padova 35122, Italy; \\	
           		$^5$ Scuola Normale Superiore, Piazza dei Cavalieri 7, Pisa 56126, Italy;\\
            		$^6$ Osservatorio Astrofisico di Arcetri, Istituto Nazionale di Astrofisica (INAF), Largo E. Fermi 5, Firenze 50125, Italy;\\
			$^7$ School of Physics, University of Melbourne, Parkville, Victoria 3010, Australia;\\
			$^8$ Osservatorio Astronomico di Padova, Istituto Nazionale di Astrofisica (INAF), Vicolo dell'Osservatorio, 5, Padova (PD) 35122, Italy.  }

   \date{Received ...; accepted ...}
 
  \abstract
{We have analyzed publicly available optical spectra of PMN J0948+0022 obtained with the Sloan Digital Sky Survey, X-Shooter, and the Multi Unit Spectroscopic Explorer (MUSE). Initially, PMN J0948+0022 was classified as a jetted narrow-line Seyfert 1 galaxy, but X-Shooter and MUSE observations, which have better spectral resolution, revealed a different profile for the H$\beta$ line, from Lorentzian to a composite one (a combination of a broad and a narrow Gaussian), more typical of intermediate Seyfert galaxies. According to the unified model, intermediate Seyferts are viewed at larger angles. However, we show that, in this case, the composite line profile results from the interaction of the powerful relativistic jet with the narrow-line region. The jet transfers part of its kinetic energy to the narrow-line region, producing flux changes in the H$\beta$ narrow component (drop by a factor of 3.4 from SDSS to X-Shooter), [O III]$\lambda$5007 core component (which nearly doubled from X-Shooter to MUSE), and its blue wing ($\Delta v\sim 200$ km s$^{-1}$), which we interpret as evidence of an outflow. We also recalculated the physical parameters of this AGN, obtaining a black hole mass of 10$^{7.8}$ M$_{\odot}$ and an Eddington ratio of $\sim$0.21 (weighted mean).}

   
   \authorrunning{B. Dalla Barba et al.}
   \titlerunning{{Interaction of the relativistic jet and the narrow-line region of PMN J0948+0022}}

   \maketitle

\section{Introduction} \label{sec1}
Classically, jetted active galactic nuclei (AGN) are divided into two main categories: radio galaxies and blazars. Blazars are further subdivided into BL Lac objects (BL Lacs) and flat-spectrum radio quasars (FSRQs), depending on the properties of their optical lines\citepads{2019ARA&A..57..467B}. Physically, FSRQs are characterized by efficient accretion and a high-density circumnuclear environment, while BL Lacs typically display inefficient accretion and a low-density circumnuclear environment. This scenario, known as the blazar sequence (\citeads{1998MNRAS.299..433F};\citeads{1998MNRAS.301..451G}), was revised when, in 2008, another class of high-energy jetted AGN was discovered: jetted narrow-line Seyfert 1 galaxies (NLS1s). Before 2008, all known NLS1s had not been detected in $\gamma-$rays, and consequently were not associated to a powerful jet emission. These objects are classified according to: broad H$\beta$ with full width at half maximum FWHM(H$\beta$) $\lesssim$ 2000 km s$^{-1}$, weak oxygen lines with [O III]/H$\beta\lesssim$ 3, and a significant iron emission (\citeads{1985ApJ...297..166O};\citeads{1989ApJ...342..224G}). Like FSRQs, jetted NLS1s are characterized by an efficient accretion process, but they exhibit relatively low black hole masses ($\sim10^7$ M$_{\odot}$), whereas both BL Lacs and FSRQs are characterized by more massive black holes ($\sim10^{8-9}$ M$_{\odot}$). There is still an ongoing debate about the actual black hole mass range for NLS1s ($\sim10^{6-8}$ M$_{\odot}$ versus $\sim10^{8-9}$ M$_{\odot}$), see for example \citeads{2020Univ....6..136F}. In this context, $\gamma-$ray emitting NLS1s represent a new class of objects that challenge the well-established theories and classifications proposed in recent years.

The first object classified as a jetted NLS1 is PMN J0948+0022 (z=0.585$\pm$0.001\footnote{Sloan Digital Sky Survey-DR9\citepads{2012ApJS..203...21A}}). Its first optical spectrum (\citeads{2002AJ....124.3042W};\citeads{2003ApJ...584..147Z}) revealed typical NLS1 characteristics: FWHM(H$\beta$) = (1500$\pm$55) km s$^{-1}$, [O III]/H$\beta$ < 3, and significant Fe II emission\citepads{2003ApJ...584..147Z}. At the same time, PMN J0948+0022 exhibited remarkable radio activity for an NLS1\citepads{2003ApJ...584..147Z}. The inverted spectrum and high brightness temperature suggested a relativistic jet closely aligned with the observer’s line of sight, which was confirmed by detections from the Fermi Gamma-ray Space Telescope (\citeads{2009ApJ...707..727A};\citeads{2009ApJ...707L.142A};\citeads{2010ASPC..427..243F}), with a jet angle of $\theta \sim 3-6^{\circ}$ (\citeads{2009ApJ...707L.142A};\citeads{2012A&A...548A.106F}). Additionally, the jets exhibit both parsec- and kiloparsec-scale structures (\citeads{2012ApJ...760...41D};\citeads{2016AJ....152...12L}), with superluminal speeds of 11.5$c$.

Between 2008 and 2016, PMN J0948+0022 was particularly active in the radio and $\gamma-$ray bands, experiencing two major outbursts. The first occurred in July 2010, reaching an isotropic luminosity of ~$10^{48}$ erg s$^{-1}$ (\citeads{2010ATel.2733....1D};\citeads{2011MNRAS.413.1671F}), and the second in January 2013, with a luminosity of ~$1.5\times10^{48}$ erg s$^{-1}$\citepads{2015MNRAS.446.2456D}. Extensive multiwavelength monitoring campaigns began shortly after the discovery and gained interest following the first flare, revealing significant variability across multiple bands (\citeads{2009ApJ...707..727A};\citeads{2011MNRAS.413.1671F};\citeads{2012A&A...548A.106F};\citeads{2014MNRAS.438.3521D};\citeads{2015A&A...575A..55A};\citeads{2017A&A...603A.100L};\citeads{2023MNRAS.523..441Y};\citeads{2022RAA....22g5001X}), a common feature of jetted AGN. These studies documented intra-night optical variability (INOV) and mid-infrared variability (\citeads{2010ApJ...715L.113L};\citeads{2021RNAAS...5..109M};\citeads{2021MNRAS.501.4110O};\citeads{2022MNRAS.514.5607O}), as well as quasi-periodic oscillations in the Fermi-Large Area Telescope energy band\citepads{2017ApJ...849...42Z}. The INOV is thought to be related to relativistic beaming\citepads{2022MNRAS.514.5607O}, while the quasi-periodic oscillations of $\sim490$ days are likely linked to jet structure\citepads{2017ApJ...849...42Z}.

The X$-$ray spectrum of PMN J0948+0022 is also variable, with a photon index $\Gamma$ ranging from 1.3 to 1.8\citepads{2015A&A...575A..13F}. The spectrum is characterized by a soft excess below 2.5 keV, possibly associated with thermal Comptonization (typical Seyfert emission), and a potential a power-law component at higher energies, likely originating from the relativistic jet (\citeads{2014MNRAS.440..106B};\citeads{2019A&A...632A.120B}).

The host galaxy of PMN J0948+0022 exhibits a brightness profile with a Sérsic index of $n<2$ for the bulge component\citepads{2020MNRAS.492.1450O}, suggesting a disk-like morphology, which is typical of NLS1 galaxies (e.g.,\citeads{2023A&A...679A..32V}).

Since the discovery of PMN J0948+0022 as a $\gamma-$NLS1, about two dozen of this type of object have been identified\citepads{2022Univ....8..587F}, suggesting that these are not rare, but instead form a significant subclass of AGN. Jetted NLS1 galaxies share several properties with other jetted AGN, including double-peaked spectral energy distributions\citepads{2009ApJ...707L.142A}, variable flat/inverted radio spectra (\citeads{2015A&A...575A..55A};\citeads{2017A&A...603A.100L}), core-jet morphology (\citeads{2006PASJ...58..829D};\citeads{2011A&A...528L..11G}), and superluminal motion\citepads{2016AJ....152...12L}. The key connection between the different families of jetted AGN likely lies in their evolutionary stage. NLS1 galaxies are thought to be young or rejuvenated quasars\citepads{2000MNRAS.314L..17M}, potentially representing the progenitors of more evolved quasars (\citeads{2016A&A...591A..88B};\citeads{2017FrASS...4....8B}).

\noindent In this paper, we analyze the optical spectral variations of the emission lines, particularly H$\beta$ and [O III]$\lambda\lambda$4959,5007, of PMN J0948+0022. We found a variation in the flux of the [O III]$\lambda$5007 core component and outflow, along with a variation in the outflow velocity. We also propose a reclassification of this AGN from a NLS1 to an intermediate Seyfert (IS), as the improved spectral resolution of X-Shooter and MUSE allowed us to disentangle the narrow and the broad components of H$\beta$. The reclassification, combined with the small jet viewing angles reported by \citeads{2009ApJ...707L.142A};\citeads{2012A&A...548A.106F};\citeads{2019MNRAS.487..640D}, makes PMN J0948+0022 a peculiar case in which the geometric interpretation of AGN provided by the unified model (UM) (\citeads{1980AJ.....85..198K};\citeads{1993ARA&A..31..473A};\citeads{1995PASP..107..803U}) raises intriguing doubts.

The paper is divided as follows. In Section~\ref{sec2} we present the optical spectral observations of PMN J0948+0022 from the Sloan Digital Sky Survey (SDSS) in 2000, X-Shooter (European Southern Observatory/Very Large Telescope/Unit Telescope 3 - ESO/VLT/UT3) in 2017, and the Multi-Unit Spectroscopic Explorer (MUSE - ESO/VLT/UT4) in 2022/2023. In Section~\ref{sec3} we present the data analysis of the mentioned spectra with a description of the fitting procedure. In Section~\ref{sec4} we show the results of the analysis: reclassification of PMN J0948+0022, variations of the H$\beta$ profile, variations of the [O III]$\lambda$5007 core and wing fluxes, estimation of the physical parameters (black hole mass, M$_{\rm BH}$, and Eddington ratio, R$_{\rm Edd}$), and calculation of the geometrical parameters (sublimation, outer, and NLR radii). In Section~\ref{sec5} we discuss the physical explanation of the decrease in flux of the H$\beta$ line and of the [O III]$\lambda$5007 line. Finally, in Section~\ref{sec6} we summarize the main results and present the conclusions.

Throughout this paper, we adopt a standard $\Lambda$CDM cosmology with: H$_{\rm{0}}$=73.3 km s$^{-1}$ Mpc$^{-1}$, $\Omega_{\rm{matter}}$=0.3, and $\Omega_{\rm{vacuum}}$=0.7\citepads{2022ApJ...934L...7R}.

\section{Observations and data reduction}\label{sec2}
\begin{figure*}
    \centering
    \includegraphics[width=1\textwidth]{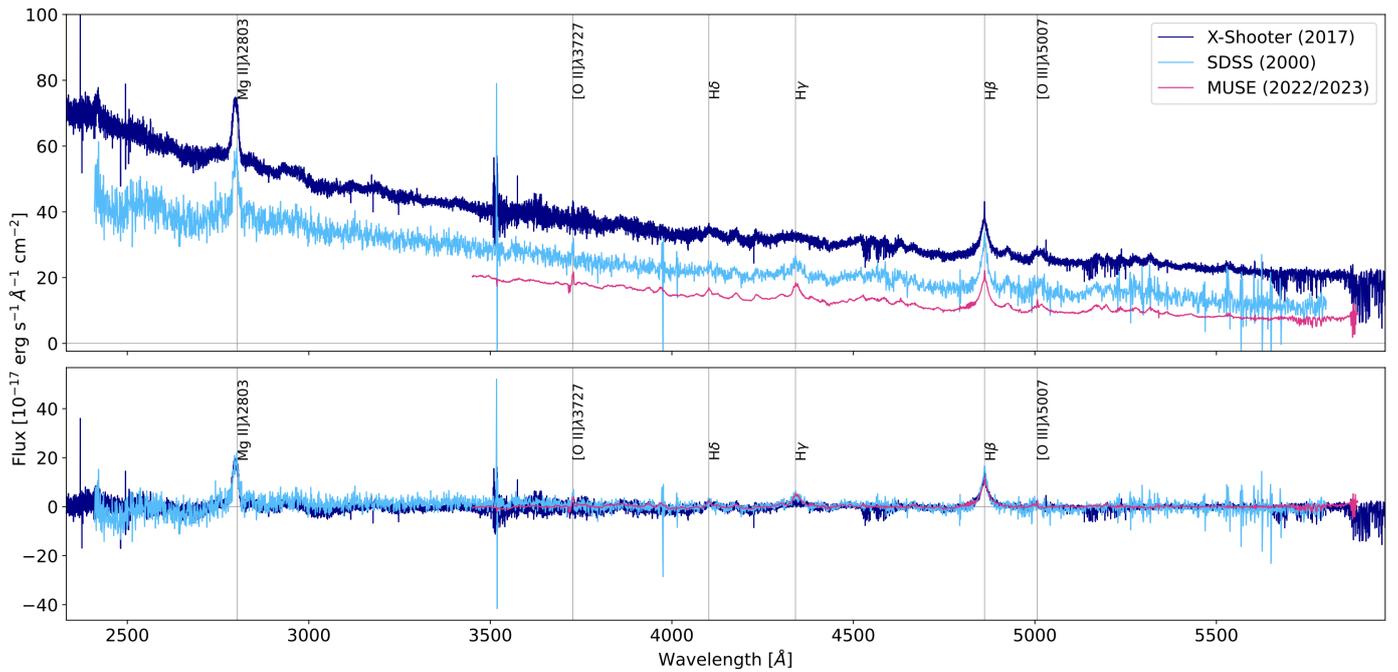} 
    \caption{At the top, the three redshift-corrected spectra in the visible range; on the bottom, the three redshift-corrected, continuum and iron lines subtracted spectra (the X-Shooter spectrum is composed of three sections, here we report the only the one useful for the comparison with the SDSS and MUSE spectra). In both panels, in light blue we present the SDSS spectrum, in blue the X-Shooter spectrum, and in magenta the MUSE spectrum. The vertical lines correspond to the identified optical lines with a significance$>5\sigma$ in all the spectra, except for [O III]$\lambda$5007 which was included even though if was not detected in the SDSS case.}
    \label{fig1}
\end{figure*}

The optical spectroscopic observations of PMN J0948+0022 can be grouped into three periods: in 2000 from SDSS, in 2017 from X-Shooter, and finally between November 24, 2022, and March 8, 2023, from MUSE. All these data are publicly available. The analysis is presented in the following sections separately, for a summary see Table~\ref{tab1}. The three redshift-corrected (z$\sim$0.5850) spectra are plotted in Fig.~\ref{fig1}.

\begin{table}
\caption{Summary of the analyzed observations of PMN J0948+0022.}
\centering
	\begin{tabular}{lllll}
            \hline
		{\bf Inst.}& {\bf R} & {\bf Obs. date} & {\bf Exp. time} & {\bf Seeing}\\
		 & & {\small\bf [dd-mm-yyyy]} &	{\small\bf [s]} & {\small\bf ["]} \\
            \hline
            \hline
		SDSS & 1500	& {\it 28-02-2000} & 3600 & -\\
		SDSS & 1500 	& 27-03-2000 & 3600  & -\\
            \hline
		X-Shooter & 6500 & 22-12-2017 & 235$\times$42 & 0.36/0.56 \\
            \hline
		MUSE & 2500	& 24-11-2022 & 1347 & 0.68 \\
			&	& {\it 25-11-2022} & 1347  & 0.86 \\
			&	& 23-01-2023 & 2559 & 0.93 \\
			&	& 28-01-2023 & 2539 & 1.37 \\
			&	& {\it 09-02-2023} & - & 1.16 \\
			&	& {\it 10-02-2023} & - & 1.19 \\
			&	& 14-02-2023 & 2514 & 1.04 \\
			&	& 08-03-2023 & 2551 & 0.54 \\
            \hline
	\end{tabular}
	\label{tab1}
	\tablefoot{The columns correspond to: instrument, resolution (as $\lambda/\Delta\lambda$), date, exposure time, and seeing conditions (for X-Shooter we present the minimum and the maximum values during the observations). The Obs. dates in italics refer to rejected observations. In particular: 28-02-2000 classified as not the optimal one from SDSS, 25-11-2022 was carried with clouds, and 09/10-02-2023 aborted, see Section~\ref{sec2.3}).}
\end{table}

\subsection{SDSS (2000)}\label{sec2.1}
PMN J0948+0022 (SDSS J094857.32+002225.5) was observed twice in 2000, the first time on February 28, 2000 (MJD 51602), and the second on March 27, 2000 (MJD 51630). The observing time was 3600s for both observations with a spectral coverage of 3800--9200\AA~and coordinates (RA,DEC)=(147.2389,0.3738) deg. The SDSS spectrograph is a multi-fiber instrument (640 fibers in total) covering a projected scale of 3 arcsec, the physical scale for this object is s=6.31 kpc/arcsec\footnote{\url{ https://astro.ucla. edu/~wright/CosmoCalc.html}}. According to\citeads{2020MNRAS.492.1450O} the effective radius for the bulge of the host galaxy of PMN J0948+0022 is $\sim$1.82/0.90 kpc (J/K band). This implies that the entire AGN is included within the SDSS aperture. The reduced spectra can be found in the SDSS SkyServer DR18\footnote{\url{https://skyserver.sdss.org/dr18/}}. Here, we used the second observation (MJD 51630) due to the SDSS automatic classification of observations (see the "sciencePrimary" label in the archive).

\subsection{X-Shooter (2017)}\label{sec2.2}
The object was observed on December 22, 2017 (MJD 58109), with 42 exposures of 235s each, for a total of approximately 2.7 hours. The spectral coverage of the instrument is divided into three arms, as the instrument itself: ultraviolet (3000--5600\AA), visible (5500--10200\AA), and near-infrared arm (10200--24800\AA). This broad spectral coverage allows us to have multiple lines in the X-Shooter spectrum (see, for example, Table~\ref{tab2}). The pointing was centered at (RA, DEC) = (147.2389, 0.3731) deg, in a slightly different position with respect to the SDSS. However, due to the distance of the object and considering the slit width (1.2''), we can conclude that the AGN was completely included in the observation (effective radius for the bulge emission of $\sim0.6/0.3"$, see again\citeads{2020MNRAS.492.1450O}). 

The data are publicly available in the ESO archive\footnote{\url{http://archive.eso.org/eso/eso_archive_main.html}}. To combine and reduce the spectra, we used the X-Shooter ESO pipeline {\it EsoReflex} version 3.5.2 (for details, see the documentation of the latest version\footnote{\url{https://ftp.eso.org/pub/dfs/pipelines/instruments/xshooter/xshoo-reflex-tutorial-3.6.8.pdf}}). The data reduction was performed using standard procedures, except for the extraction radius of the one-dimensional spectrum obtained from the two-dimensional exposure. A smaller extraction region was selected to minimize potential contamination from the host galaxy. For this purpose, we defined the extraction region based on the observing seeing conditions on December 22, 2017, as the observations were seeing-limited. During that night, the seeing varied between 0.56'' and 0.36'' (see Table~\ref{tab1}). To be conservative, we adopted 0.6'' (FWHM) as a reference and converted it into pixels using a scale of 0.158''/px (see the spectrograph characteristics\footnote{\url{https://www.eso.org/sci/facilities/paranal/instruments/xshooter/inst.html}}). The final extraction radius is 5px (compared the default one of the pipeline, which is 15px). The extraction procedure was conducted using {\tt IRAF version 2.18}. Finally, we obtained three FITS files, corresponding to the three arms of the spectrograph: NIR, visible, and UV. 

\subsection{MUSE (2022/2023)}\label{sec2.3}
MUSE is an integral field spectrograph covering the spectral range 4650--9300 \AA, producing data cubes with a 1×1 arcmin$^2$ field of view (FoV) in wide field mode (WFM). These data cubes have two spatial dimensions (x, y) and one spectrum per pixel (see Fig.~\ref{fig_a2} in Appendix C), where the spectral information is stored along the z-axis. MUSE observations of PMN J0948+0022 were carried in the WFM without adaptive optics (NOAO). The object was observed multiple times and under different conditions; for details, see Table~\ref{tab1}. We rejected from the analysis the observations of November 25 and February 9-10 from the analysis due to poor data quality, mainly related to the atmospheric conditions (observations conducted with clouds, or aborted for the same reason). The reduction of the spectra, similarly to the X-Shooter case, was carried out with the MUSE ESO pipeline ({\it EsoReflex}) version 2.8.5. The final products are single reduced exposures and the combined datacube. Before the spectra extraction, we applied {\tt CubePCA}{\footnote{\url{https://github.com/brandherd/CubePCA}}}, a package developed for cleaning astronomical IFU observations -- particularly MUSE data cubes -- from sky line residuals and systematic artifacts. The core concept of {\tt CubePCA} is to select a reference region in the cube (top left in our case) to represent the sky, create a mask, and apply Principal Component Analysis (PCA) to subtract the sky contribution from the entire cube. After this cleaning step, we extracted the spectrum from a circular region with a radius ranging from 4 to 9 pixels, centered at (RA, DEC) = (147.2388, 0.3738) deg. This differential extraction allowed us to take into account for the different weather conditions, using larger extraction radii for the worst seeing conditions (such as those on January 28, 2023) and smaller radii for the best cases (such as those on March 8, 2023), see Table~\ref{tab1}. Given the point-like nature of the source, we converted each seeing (interpreted as a FWHM) into pixels to ensure that only the AGN contribution was included in the extracted spectrum. The pixel scale of 0.2"/px confirms that, in all cases, the observations are seeing-limited. We then inspected the individual MUSE spectra and compared the H$\beta$--[O III]$\lambda$5007 profiles. From individual spectra, we confirmed that the line fluxes do not vary significantly ($\lesssim2\sigma$) during the time spanned by the observations. Keeping in mind this, we combined the individual MUSE spectra using a weighted average, where the weights were determined by the continuum fluctuations in the 5050--5150\AA~range. This procedure allows us to reduce the uncertainties of the flux.

\section{Data analysis}\label{sec3}
\subsection{Methodology}\label{sec3.1}
We performed line identification in the PMN J0948+0022 spectra. We set the significance level at 5$\sigma$, only the lines that met this threshold were used to calculate the weighted mean for the redshift (see Table~\ref{tab_a1}). Here, $\sigma$ is the standard deviation of the line-free continuum in the 5050--5150\AA~range. For [O II]$\lambda$3727, H$\beta$, [O III]$\lambda$5007, H$\alpha$, and [N II]$\lambda$6583 (when present) we performed an accurate fitting, described in Section~\ref{sec3.2}, the results of which are reported in Table~\ref{tab2} where we listed: the identified lines with the observed wavelength ($\lambda_{\rm{obs}}$), the flux, and the FWHM. 

We corrected the wavelength by the redshift using the [O II]$\lambda$3727 line, which is the [O II]$\lambda$3726,3729 doublet in the X-Shooter and MUSE cases, and a single line in the SDSS case. The difference is due to the limited spectral resolution of the instrument. We chose [O II]$\lambda$3727 because of its presence in all the spectra and its significance with respect to [O III]$\lambda$5007. From its displacement, we obtained: $z=(0.5857\pm0.0001)$ in SDSS, $z=(0.5850\pm0.0001)$ in X-Shooter, and $z=(0.5850\pm0.0004)$ in MUSE. Each value is in agreement with the public value of $z=0.585$.

For the modeling of the continuum, host galaxy, and iron components, we used AGN FANTASY (Fully Automated pythoN tool for AGN Spectra analYsis)\footnote{\url{https://fantasy-agn.readthedocs.io/en/latest/}} (\citeads{2020A&A...638A..13I};\citeads{2022MNRAS.516.1624R};\citeads{2023ApJS..267...19I}). This tool is based on the Levenberg-Marquardt algorithm, implemented through the {\tt sherpa} Python package, and optimized for simultaneous multi-component fitting of AGN spectra. The continuum was found to be AGN-dominated and was fitted with a power law with a spectral index of $\sim$3. The host-galaxy contribution was negligible. For the modeling of the iron lines, we used the AGN FANTASY template obtained for the MUSE spectrum, and then subtracted it to the other two cases. 

Finally, it is worth mentioning that the original X-Shooter spectrum exhibited significant telluric contamination in the H$\beta$ line, which we discussed this point in Appendix A (see Fig.~\ref{fig_a1}). 

\subsection{Line fitting}\label{sec3.2}
We performed the line fitting using a Python (version 3.12) code, based on the {\tt scipy.optimize.curve\_fit} module. The module allows the user to set the initial parameters of the fitting, define the fitting curve, and also set constrains on the parameters. In the following we define the used functions, and the eventual bounds between the parameters (e.g. relations between the FWHM of the components, or theoretical ratios of lines).

The first step was to identify a significant narrow line to tie to the narrow component of the permitted lines. While [O III]$\lambda\lambda$4959,5007 is typically the preferred choice, in PMN J0948+0022, the core component of these oxygen lines ([O III]$\lambda5007_{\rm c}$) was too faint to be useful. Consequently, we selected the [O II]$\lambda$3727 line. This choice was driven by two factors: first, this line is detected in all three spectra (due to the spectral coverage of the instruments), and second, it is detected with a greater significance compared to [O III]$\lambda$5007 (see Table~\ref{tab2}). This allows us to apply the same procedure to all the cases. The [O II]$\lambda$3727 line is actually a doublet ([O II]$\lambda\lambda$3726,3729), but in the SDSS spectrum, the spectral resolution is insufficient to separate the two lines, so we treated it as a single line. In the X-Shooter and in the MUSE spectra, however, the two lines can be distinguished, so we performed the fitting of both of them and used the averaged FWHM as a reference for the other forbidden lines. 

As usual, we also linked the amplitude of the two [O III]$\lambda\lambda$4959,5007 lines to the theoretical ratio of (2.993$\pm$0.014) (\citeads{2006agna.book.....O};\citeads{2007AIPC..895..313D}). The same applies to the [N II]$\lambda\lambda$6548,6583 lines, only visible in the X-Shooter spectrum due to its larger spectral coverage. Here the theoretical value of the ratio of the lines is (3.039$\pm$0.021)\citepads{2023AdSpR..71.1219D}.

All the lines were fitted with a single or a combination of Gaussian profiles. For the forbidden lines, a single profile was sufficient to reproduce the emission, except in the case of [O III]$\lambda$5007. In this line, a significant blue wing is present, which was fitted with an additional Gaussian ([O III]$\lambda5007_{\rm o}$). In the SDSS case, the wing is mixed with the noise continuum (< 3$\sigma$), so we adopted the same model as in the X-Shooter and MUSE cases but considered the [O III]$\lambda$5007 line in SDSS as an upper limit.	

For the permitted lines, we attempted to fit all three cases with a single Lorentzian profile as done by\citeads{2003ApJ...584..147Z}, but this model yielded a good result in terms of reduced chi squared ($\chi^2_{\rm red}$) only in the SDSS case ($\chi^2_{\rm{red}}\sim1.6$). Specifically, for the X-Shooter and MUSE cases, this initial model could not reproduce the narrow component (H$\beta_{\rm n}$). As a second approach, we tried using two Gaussians, one for the narrow component and one for the entire broad component (H$\beta_{\rm b}$), but this result was also unsatisfactory. In the X-Shooter and MUSE cases, the best $\chi^2_{\rm red}$ was achieved by using a combination of three Gaussians: one for the narrow component and two for the broad component (H$\beta_{\rm b1}$ and H$\beta_{\rm b2}$, their combination corresponds to the entire broad component H$\beta_{\rm b}$ of Table~\ref{tab2}). Physically, the adopted model should be interpreted as a composition between the emission from the NLR and a stratified BLR (e.g.,\citeads{2006SerAJ.173....1P}). Finally, to ensure consistency in comparing results, we adopted this three-Gaussian model for the SDSS case.

\begin{figure}
    \centering
    \includegraphics[width=0.45\textwidth]{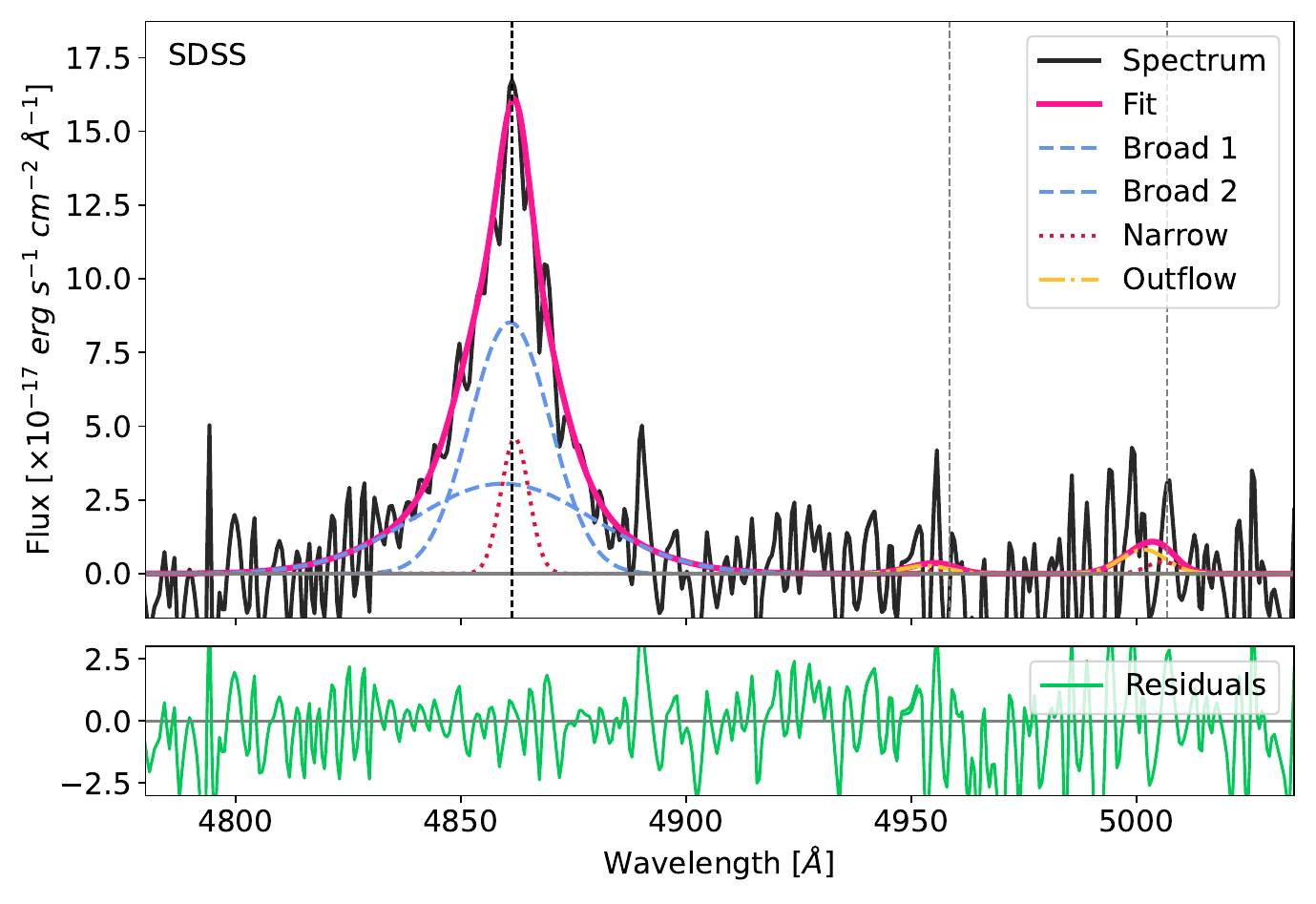} 
    \includegraphics[width=0.45\textwidth]{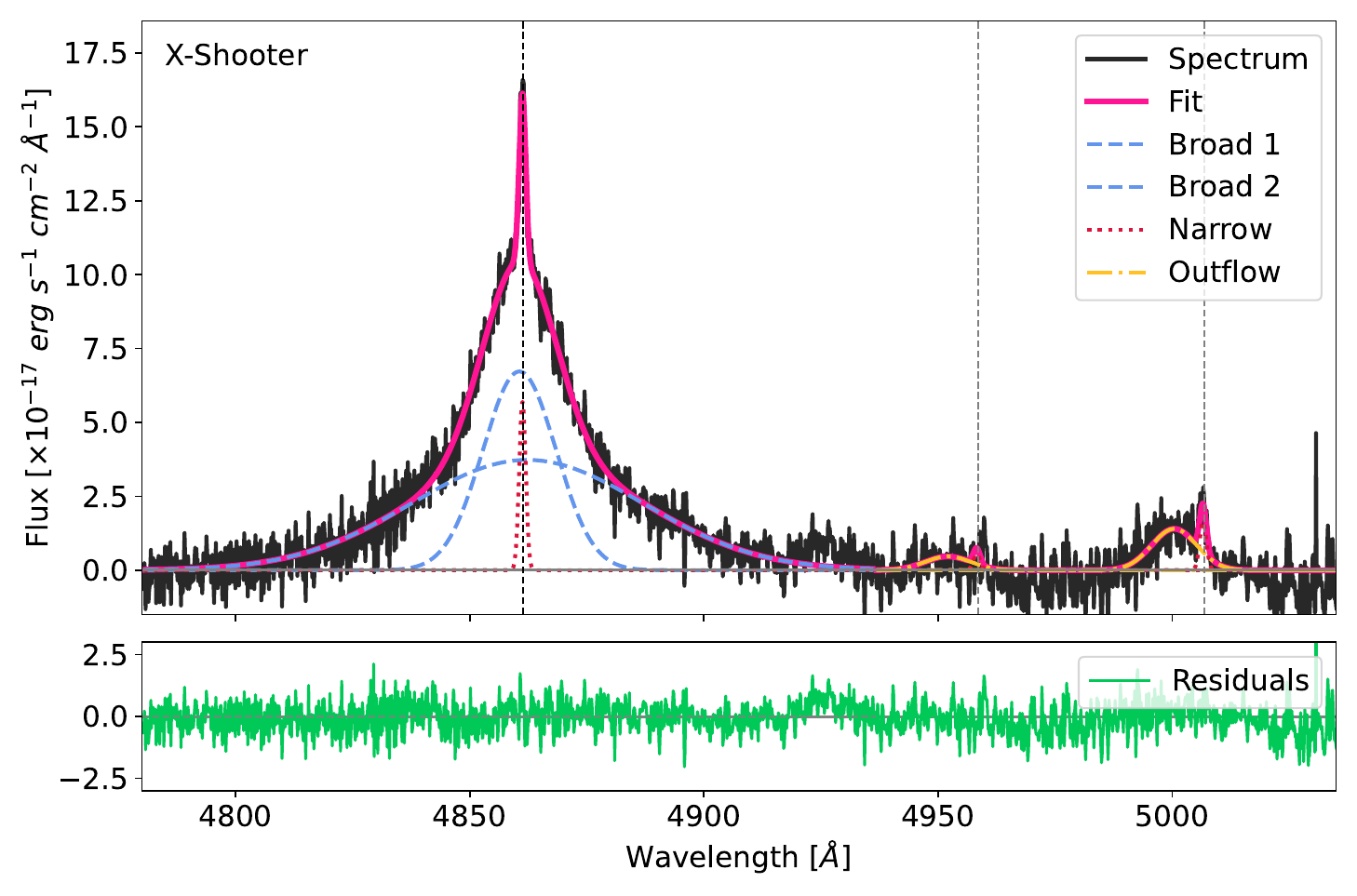} 
    \includegraphics[width=0.45\textwidth]{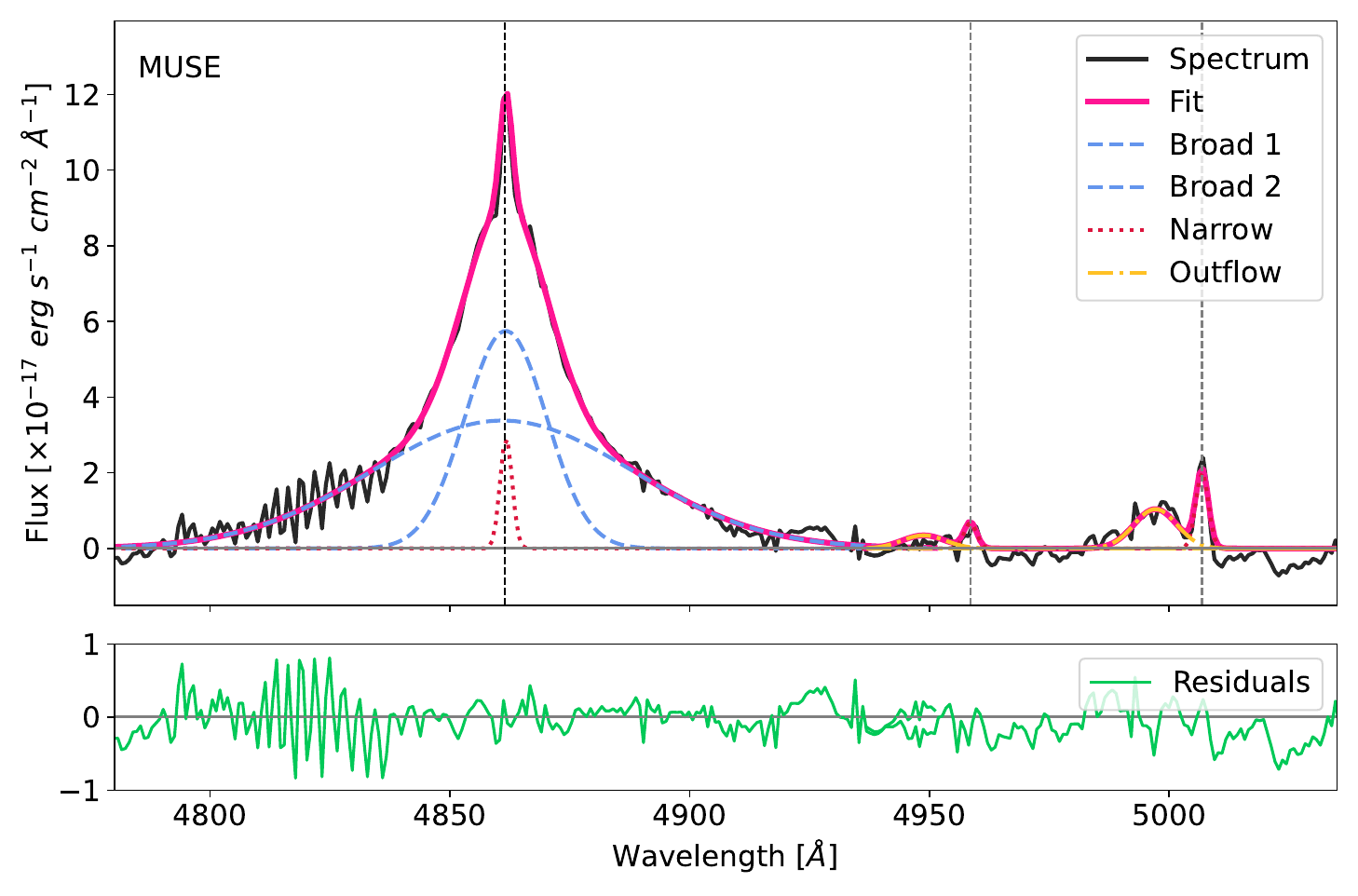} 
    \caption{H$\beta$--[O III]$\lambda\lambda$4959,5007 spectral region in the SDSS (top panel), X-Shooter (middle panel), and MUSE (bottom panel) spectra of PMN J0948+0022. Colors: in magenta continuum line the final fit, in blue dashed lines the broad components of H$\beta$, in red dotted lines the narrow components of H$\beta$ and [O III]$\lambda\lambda$4959,5007, in yellow dotted-dashed lines the [O III]$\lambda\lambda$4959,5007 blue wings interpreted as outflows, and finally in green lines the residuals calculated as the difference between the original data and the best-fitting model. For SDSS the [O III]$\lambda$5007 fitting is tentative; here, we applied the same model as in the other two cases, keeping in mind that we are dealing with a non-detected line.}
    \label{fig2}
\end{figure}

The H$\beta$--[O III]$\lambda\lambda$4959,5007 regions are plotted in Fig.~\ref{fig2}. All the fitting parameters can be found in Tables~\ref{tab2} and~\ref{tab_a1}. The errors in the parameters were calculated through a Monte Carlo method, creating N=1000 mock spectra and using the continuum fluctuations at 5100$\AA$. In the X-Shooter and MUSE cases we also added a 5\% in the parameter errors to take into account for the flux calibration systematics introduced by the pipeline reduction (\citeads{2014A&A...572A..13S};\citeads{2020A&A...641A..28W}). For the derived quantities, the errors are calculated using standard error propagation.

\subsection{The resolution problem}\label{sec3.3}
\begin{figure}
    \centering
    \includegraphics[width=0.45\textwidth]{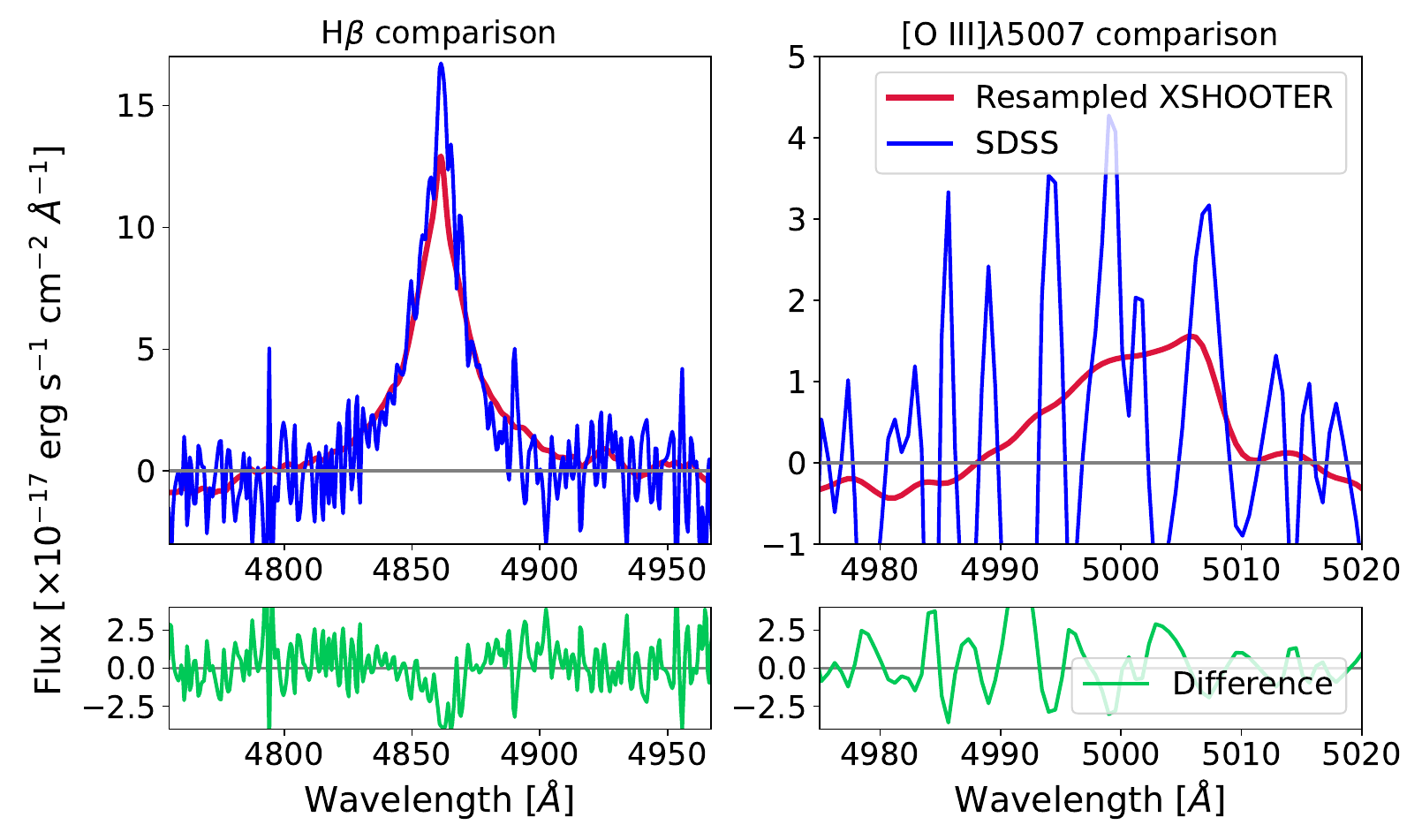} 
    \includegraphics[width=0.45\textwidth]{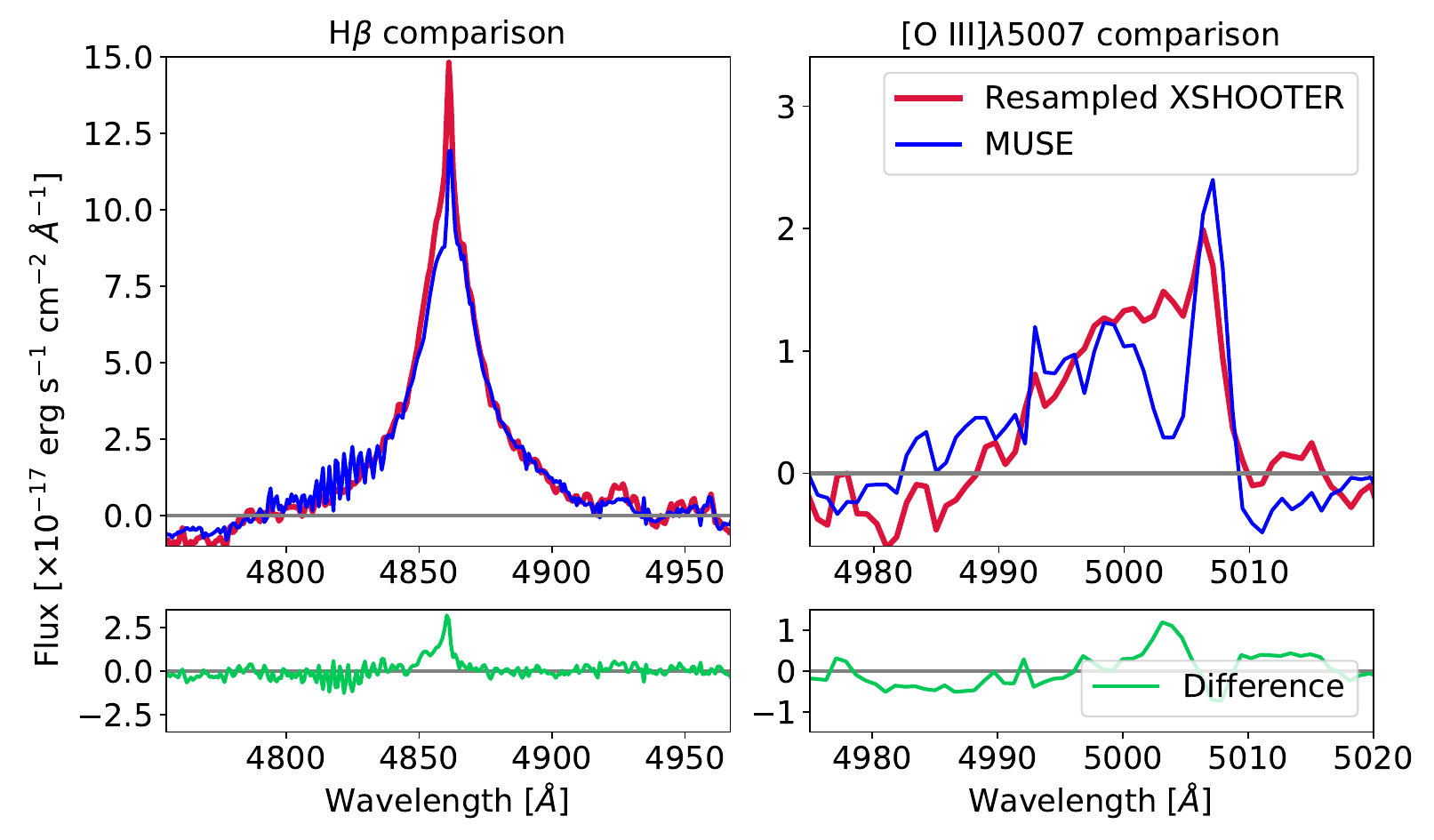} 
    \caption{(Top panels) Comparison of the H$\beta$ and [O III]$\lambda$5007 (left and right panels, respectively) lines observed in the SDSS$-$X-Shooter spectra with the resolution of the latter decreased to match that of the former (SDSS in blue and X-Shooter in red). (Bottom panels) The same as in the top panels, but for X-Shooter $-$ MUSE with the resolution of the former reduced to match the that of the latter (MUSE in blue and X-Shooter in red). The left panels refer to the comparison of H$\beta$, while the right panels to [O III]$\lambda$5007.}
    \label{fig3}
\end{figure}

When comparing spectra from different instruments, it is important to verify whether the line variability can be attributed to the different instrumental resolutions. X-Shooter has the best spectral resolution (R$=\lambda / \Delta\lambda\sim6500$), followed by MUSE (R$\sim$2500), and SDSS (R$\sim$1500). We investigated the possibility that the variations in the lines were related solely to spectral resolution differences. To do so, we lowered the resolution of the X-Shooter spectrum, to match those of SDSS and MUSE. Specifically, we convolved the line profiles with the line spread function expressed in terms of the R parameter (first approximation). The results are plotted in Fig.~\ref{fig3}. 

In the top-left panel of Fig.~\ref{fig3}, we can find the difference between X-Shooter and SDSS in the H$\beta$ region, featuring a reduced resolution for the first spectrum. The difference is mainly in the central part, with a lower flux for the X-Shooter case. We note that while the H$\beta$ profile in the SDSS spectrum can be fitted with a simple Lorentzian profile, this is no longer suitable for X-Shooter due of the prominence of H$\beta$ narrow component. This corresponds to the typical spectral profile of an IS (e.g.,\citeads{1943ApJ....97...28S};\citeads{1976MNRAS.176P..61O}). We note that the flux of H$\beta_{\rm n}$ as measured by SDSS is more than three times that of X-Shooter. Nevertheless, the low spectral resolution of SDSS hampers a complete disentangling of the different components of H$\beta$. Therefore we conclude that the original classification as NLS1 was biased by the lower spectral resolution of SDSS.

In the bottom-left panel, we can see that H$\beta$ in the MUSE case exhibits a lower flux in the central part compared to the X-Shooter spectrum.
From the results in Table~\ref{tab2} (fitting of the unconvolved spectra), we observe that the change in the H$\beta_{\rm n}$ between X-Shooter and MUSE is weakly significant ($\sim$14\%, $\sim 3 \sigma$).

In the top-right panel, we can find the same figure as the left panel, but for [O III]$\lambda$5007. Given that the measured upper limit in SDSS does not constrain the flux, it is not possible to draw important conclusions regarding its comparison with the resolution-decreased X-Shooter spectrum. In the other case, we reported a significant change (> 8$\sigma$) of the core component of [O III]$\lambda$5007 (flux almost doubled), while the blue wing is more asymmetric in the MUSE profile (see Fig.~\ref{fig3} bottom-right panel) with the MUSE flux equal to about 71\% of the X-Shooter flux.

\section{Results}\label{sec4}

\subsection{H$\beta$ profile}\label{sec4.1}
The flux of the H$\beta$ total broad component remains stable across all the observations (variations within 1$\sigma$), as shown in Table~\ref{tab2}. However the profile is changing as indicated by the relative change of the two sub-broad components (broad 1 and 2). H$\beta_{\rm b1}$ is stable in X-Shooter and MUSE ($\sim230\cdot 10^{-17}$ erg s$^{-1}$cm$^{-2}$ both), but it shows a lower flux in SDSS ($\sim154\cdot 10^{-17}$ erg s$^{-1}$cm$^{-2}$), while H$\beta_{\rm b2}$ changes in all the three cases (SDSS $\sim190\cdot 10^{-17}$ erg s$^{-1}$cm$^{-2}$, X-Shooter$\sim132\cdot 10^{-17}$ erg s$^{-1}$cm$^{-2}$, and MUSE $\sim122\cdot 10^{-17}$ erg s$^{-1}$cm$^{-2}$). As we noted in Section~\ref{sec3.2} the two broad components can be interpreted as the emission from a stratified BLR\citepads{2006SerAJ.173....1P}, therefore changes in H$\beta_{\rm b1}$ and H$\beta_{\rm b2}$ can be interpreted as changes in the radial profile of the BLR. In addition,\citeads{2016MNRAS.462.1256C} noted that this profile is more typical of broad-line Seyfert 1 (BLS1), which strengthens our reclassification of PMN J0948+0022 as IS. 

H$\beta_{\rm n}$ drastically decreased from SDSS to X-Shooter epochs (dropping by a factor of 3.4, about 4$\sigma$), while the change from the latter to MUSE is smaller (MUSE flux is 86\% of X-Shooter one, $\sim3\sigma$), as already noted in Section~\ref{sec3.3}.

From these variations, we can estimate the halving-doubling timescale $\tau$ by using this simple formula:

\begin{equation}
	\frac{F(t_{\rm{2}})}{F(t_{\rm{1}})} = 2^{-\frac{(t_{\rm{2}}-t_{\rm{1}})}{\tau}}
	\label{eq1}
\end{equation}

\noindent where $t_{\rm{1}}$ and $t_{\rm{2}}$ denote the epochs of the observations (SDSS MJD 51630, X-Shooter MJD 58109, and MUSE MJD 59959$\pm$52, this being the middle epoch of all the MUSE observations), F$(t_{\rm{1}})$ and F$(t_{\rm{2}})$ are the fluxes to be considered, and $\tau$ represents the timescale over which the flux decreases to half (or increases to twice) its initial value.

In the H$\beta_{\rm n}$ case we obtain: $\tau\sim10_{-1}^{+2}$ years for SDSS and X-Shooter, and $\tau\sim27_{-10}^{+32}$ years for X-Shooter and MUSE (see Table~\ref{tab2} for the flux values).

\subsection{[O III]$\lambda$5007 outflow}\label{sec4.2}
The first optical spectrum of PMN J0948+0022, obtained from SDSS, was analyzed by\citeads{2003ApJ...584..147Z} who noted that the [O III]$\lambda$5007 line was not significantly detected in this spectrum. More recently,\citeads{2016A&A...591A..88B} examined the [O III] properties of $\gamma-$ray emitters identified by\citeads{2015A&A...575A..13F}, classifying PMN J0948+0022 among the blue outliers. These are objects with strong [O III] blueshifts exceeding 100 to several hundred km s$^{-1}$ (e.g.,\citeads{2008ApJ...680..926K}). The classification by\citeads{2016A&A...591A..88B} was driven by the presence of the prominent blue wing interpreted as an outflow, rather than the oxygen core component which is completely buried in the noise in the SDSS spectrum, as depicted in Fig.~\ref{fig2}. With the new X-Shooter and MUSE observations, which show a significant detection of this line, we can confirm PMN J0948+0022 as a blue outlier (see Tables~\ref{tab2} and \ref{tab_a1}). 

As discussed in Section~\ref{sec3}, we confirmed the lack of a reliable detection of the [O III]$\lambda$5007 line in the SDSS spectrum (significance $<3\sigma$). Consequently, we tested the best-fit model -- a narrow Gaussian for the core component and a broad blueshifted Gaussian for the outflow -- on the X-Shooter and MUSE spectra, and applied the same model to the SDSS data. Details of the parameters are provided in Tables~\ref{tab2} and \ref{tab_a1}. The results show a blue wing with a negative velocity of < 466 km/s, (383$\pm$31) km/s, and (582$\pm$20) km/s, respectively in the three epochs with an increased asymmetry in the X-Shooter and MUSE, as shown in Fig.~\ref{fig3} bottom-right panel. This confirms the classification of PMN J0948+0022 in the blue outliers family.

Comparing the different observations, we can calculate the halving-doubling timescale using Equation~\ref{eq1}. We considered only the difference between X-Shooter of MUSE, because of the non-detection of [O III]$\lambda$5007 in the SDSS spectrum. The results are: $\tau=4.8^{+1.1}_{-0.7}$ years for [O III]$\lambda$5007$_{\rm c}$, and $\tau=10^{+15}_{-4}$ years in the case of [O III]$\lambda$5007$_{\rm o}$ (see Table~\ref{tab2} for the flux values).

\subsection{Black hole mass}\label{sec4.3}
To estimate M$_{\rm{BH}}$, we used the classical virial relation:

\begin{equation}
	M_{\rm{BH}} = f \frac{R_{\rm{BLR}} \cdot \sigma_{\rm line}^2}{G}
	\label{eq2}
\end{equation}

\noindent where $f$ is a factor related to the geometry of the system, $R_{\rm{BLR}}$ is the radius of the BLR, $\sigma_{\rm line}$ is the second order moment of the line profile, and $G$ is the gravitational constant. We adopted $f=(3.85 \pm 1.15)$ as reported by\citeads{2006A&A...456...75C}. The second order moment $\sigma_{\rm line}^2(\lambda)$ is calculated as follows (see Equation~\ref{eq3}):

\begin{equation}
	\sigma_{\rm line}^2(\lambda) = \frac{\int \lambda^2 P(\lambda) \, d\lambda}{\int P(\lambda) \, d\lambda} - \left(\frac{\int \lambda P(\lambda) \, d\lambda}{\int P(\lambda) \, d\lambda}\right)^2
	\label{eq3}
\end{equation}

\noindent where $P(\lambda)$ is the profile of the selected line and $\lambda$ is its wavelength.

To estimate $R_{\rm{BLR}}$, we employed several methods. First, we used the properties of H$\beta$ and Equation~\ref{eq4} from\citeads{2010ApJ...723..409G}, revised by\citeads{2021A&A...654A.125B}:

\begin{equation}
	\mathrm{log} \frac{R_{\rm{BLR}}}{10~\mathrm{lt-day}} = (0.53 \pm 0.04) \times \mathrm{log} \frac{L(\mathrm{H}\beta_{\rm b})}{10^{43}~\mathrm{erg/s}} + (1.85 \pm 0.05)
	\label{eq4}
\end{equation}

\noindent where $L(\mathrm{H}\beta_{\rm b})$ is the luminosity of the broad component of the H$\beta$ line.

Second, we used the continuum luminosity at 5100 \AA, applying Equation~\ref{eq5} from\citeads{2013ApJ...767..149B}:

\begin{equation}
	\mathrm{log} \frac{R_{\rm{BLR}}}{1~\mathrm{lt-day}} = (0.53 \pm 0.03) \times \mathrm{log} \frac{\lambda L_{\rm{\lambda}}(5100)}{10^{44}~\mathrm{erg/s}} + (1.53 \pm 0.03)
	\label{eq5}
\end{equation}

\noindent where $\lambda L_{\rm{\lambda}}(5100)$ is the continuum luminosity at 5100\AA. 

Finally, we considered an updated method that includes the iron contribution, as suggested by\citeads{2019ApJ...886...42D}, using Equation~\ref{eq6}:

\begin{equation}
	\begin{split}
		\mathrm{log} \frac{R_{\rm{BLR}}}{1~\mathrm{lt-day}} = (0.45 \pm 0.03) & \times \mathrm{log} \frac{\lambda L_{\rm{\lambda}}(5100)}{10^{44}~\mathrm{erg/s}} + (1.65 \pm 0.06) + \\
		& + (-0.35 \pm 0.08) \times R_{4570}
		\label{eq6}
	\end{split}
\end{equation}

\noindent where $R_{4570}=F(\mathrm{Fe II}\lambda4570)/F(\mathrm{H}\beta_{\mathrm{b}})$, $F(\mathrm{Fe II}\lambda4570)$ is the flux of the iron bump between 4434--4684\AA, and $F(\mathrm{H}\beta_{\rm b})$ is the flux of the broad component of H$\beta$. 

The results are summarized in Table~\ref{tab3}. All methods yield consistent values within the uncertainties and align with the estimates presented in\citeads{2015A&A...575A..13F}. However, a noticeable bias is observed in the second method relative to the other two, indicating that the\citeads{2013ApJ...767..149B} equation may overestimate M$_{\rm{BH}}$. This method relies solely on the continuum luminosity, and since PMN J0948+0022 is a jetted AGN\citepads{2009ApJ...707L.142A}, the continuum includes contributions from both the accretion disk and the jet, which could explain the overestimation. Indeed, the jet component increases the continuum flux, leading to an overestimation of the BLR radius. 

It is also worth noting that the method incorporating the iron contribution (Equation~\ref{eq6}) may be affected by systematic uncertainties due the complexities of fitting and subtracting the iron emission. For this reason, we adopt the black hole mass estimate obtained from the first method (Equation~\ref{eq4}) by\citeads{2010ApJ...723..409G} for the subsequent analysis. We calculated the weighted average between the three results (SDSS, X-Shooter, and MUSE), obtaining log(M$_{{\rm BH}}$)=(7.76$\pm$0.12) M$_{\odot}$ (last row of Table~\ref{tab3}).

\subsection{Eddington ratio}\label{sec4.4}
Given the M$_{\rm{BH}}$ parameter, we can derive R$_{\rm Edd}$ using Equation~\ref{eq7}:

\begin{equation}
    R_{\rm{Edd}} = \frac{L_{\rm{bol}}}{L_{\rm{Edd}}} = \frac{L_{\rm{bol}}}{1.3 \times 10^{38}~M_{\rm{BH}}/M_{\odot}~\mathrm{erg/s}}=\dot{m}
    \label{eq7}
\end{equation}

where $L_{\rm{bol}}$ denotes the bolometric luminosity, which primarily reflects the accretion disk emission, and $\dot{m}$ is the accretion parameter (the last equality is true if the efficiency $\eta$ is constant). The most common method for estimating the disk luminosity is through the equation provided by\citeads{2000ApJ...533..631K}, which relates L$_{\rm{bol}}$ to the continuum luminosity at 5100\AA:

\begin{equation}
    L_{\rm{bol}} = 9 \times \lambda L_{\rm{\lambda}}(5100)
    \label{eq8}
\end{equation}

The continuum luminosity can be directly measured from the spectrum and substituted it into the above equation. Alternatively, we can invert Equations~\ref{eq5} and~\ref{eq6} to calculate the continuum luminosity from R$_{\rm{BLR}}$ obtained using Equation~\ref{eq4}. This latter approach derives the continuum based on line properties, specifically as a function of L(H$\beta_{\rm{b}})$. The inverted forms of Equations~\ref{eq5} and~\ref{eq6} are:

\begin{equation}
    \mathrm{log}\frac{\lambda L_{\rm{\lambda}}(5100)}{10^{44}~\mathrm{erg/s}} = \left[\log\left(\frac{R_{\rm{BLR}}}{1~\mathrm{lt-day}}\right) - (1.53 \pm 0.03)\right] \cdot \frac{1}{(0.53 \pm 0.03)}
    \label{eq9}
\end{equation}

\begin{equation}
\begin{split}
    \mathrm{log}\frac{\lambda L_{\rm{\lambda}}(5100)}{10^{44}~\mathrm{erg/s}} = & \left[\log\left(\frac{R_{\rm{BLR}}}{1~\mathrm{lt-day}}\right) - (1.65 \pm 0.06) \right. +\\
    & \left. + (0.35 \pm 0.08) \times R_{4570} \vphantom{\frac{R_{\rm{BLR}}}{1~\mathrm{lt-day}}} \right] \cdot \frac{1}{(0.45 \pm 0.03)}
    \label{eq10}
\end{split}
\end{equation}

All three methods are affected by different sources of bias. Directly using Equation~\ref{eq8} encounters similar issues as Equation~\ref{eq5}, as it relies on continuum measurements from the spectrum. As discussed in Section~\ref{sec4.3}, for jetted sources, the continuum includes contributions from both the accretion disk and the jet, leading to contamination in the measurements. Conversely, the other two methods derive the continuum from line properties, in particular H$\beta$. Equation~\ref{eq10} is influenced by the iron fitting process, making the second method (inverting Equation~\ref{eq5} into Equation~\ref{eq9} and using R$_{\rm{BLR}}$ from Equation~\ref{eq4}) a less biased approach in our specific scenario.

The results are shown in Table~\ref{tab3}. It is evident from the table that the second method yields the most consistent results, whereas the other two methods exhibit more variability, likely influenced by jet activity. Consequently, we used the value calculated with Equation~\ref{eq9} (central section, last column of Table~\ref{tab3}) as the reference for $\lambda L_{\lambda}(5100)$. The weighted average between the X-Shooter and MUSE values of the second method the reference value for the discussion (R$_{\rm Edd}$=0.21$\pm$0.06), as shown in the last row of Table~\ref{tab3}. 

\subsection{Geometry of the system}\label{sec4.5}
We derived the key geometrical parameters necessary to construct a schematic representation of the AGN. These include the sublimation radius ($R_{\rm{sub}}$), which represents the boundary between the BLR and the dusty torus; the outer radius (R$_{\rm{out}}$), the outer boundary of the dusty torus; and the maximum extent of the NLR (R$_{\rm{NLR}}$). These quantities were computed using the relations from\citeads{2008NewAR..52..274E} and\citeads{2018ApJ...856..102F}:

\begin{equation}
    R_{\rm{sub}} = 0.4 \sqrt{\frac{L_{\rm{bol}}}{10^{45}~\mathrm{erg/s}}}
    \label{eq11}
\end{equation}

\begin{equation}
    R_{\rm{out}} = 12 \sqrt{\frac{L_{\rm{bol}}}{10^{45}~\mathrm{erg/s}}}
    \label{eq12}
\end{equation}

\begin{equation}
    R_{\rm{NLR}} = 10^{-18.41 + 0.52 \cdot \mathrm{log}~L_{\mathrm{[O~III]}}}
    \label{eq13}
\end{equation}

All these quantities are functions of L$_{\rm{bol}}$ (reported in Table~\ref{tab3}) or of the [O III]$\lambda$5007 luminosity L$_{\mathrm{[O~III]}}$. Results are reported in Table~\ref{tab4} in parsecs. There is general agreement in the R$_{\rm{sub}}$ and R$_{\rm{out}}$ parameters, while the NLR in the SDSS case appears to be more extended. It is important to note that the NLR size is determined from the [O III]$\lambda$5007 properties (see Equation~\ref{eq13}), and in the SDSS spectrum, this line was not significantly detected, representing only an upper limit.

\begin{table}
\caption{M$_{\rm{BH}}$, R$_{\rm{Edd}}$, and L$_{\rm bol}$ along with their associated uncertainties for SDSS, X-Shooter, and MUSE spectra of PMN J0948+0022.}
\centering
\small
	\begin{tabular}{l|ll|ll|l}
            \hline
		 {\bf Spectrum}  &  {\bf Eq.} & {\bf log(M$_{\rm{BH}}$)} & {\bf Eq.} & {\bf R$_{\rm{Edd}}$} & {\bf log(L$_{\rm bol}$)} \\
		 			&		&	{\bf [M$_{\odot}$]} 	& 		&				& {\bf [erg s$^{-1}$]} \\
            \hline
            \hline
		 SDSS 		&				& 7.47 $\pm$ 0.66 	&				& <6				&	 	45.93 $\pm$ 1.03\\
		X-Shooter 	&	(\ref{eq4})		& 7.74 $\pm$ 0.14 	&	(\ref{eq8})		& 2.01 $\pm$ 0.66 	&	 	46.16 $\pm$ 1.43\\
		 MUSE 		&				& 7.83 $\pm$ 0.22 	&				& 0.63 $\pm$ 0.31 	&		45.74 $\pm$ 2.11\\
            \hline
		SDSS 		&				& 7.86 $\pm$ 0.66 	&				& <1				&	 	45.20 $\pm$ 1.02 \\
		X-Shooter 	&	(\ref{eq5})		& 8.23 $\pm$ 0.14 	&	(\ref{eq9})		& 0.23 $\pm$ 0.08 	&	 	45.21 $\pm$ 1.01\\
		MUSE 		&				& 8.11 $\pm$ 0.21 	&				& 0.18 $\pm$ 0.09 	&	 	45.20 $\pm$ 1.02\\
            \hline
		SDSS 		&				& 7.60 $\pm$ 0.77 	&				& <0.6 			&	 	44.96 $\pm$ 0.39\\
		X-Shooter 	&	(\ref{eq6})		& 7.72 $\pm$ 0.31 	&	(\ref{eq10})	& 0.59 $\pm$ 0.32 	&	 	45.62 $\pm$ 0.36\\
		 MUSE 		&				& 7.81 $\pm$ 0.22 	&				& 0.11 $\pm$ 0.07 	&	 	44.96 $\pm$ 0.39\\
	    \hline
	    \hline
	     Reference 		&				& 7.76 $\pm$ 0.12 	&				& 0.21 $\pm$ 0.06 	&		45.23 $\pm$ 0.59\\
            \hline
	\end{tabular}
	\label{tab3}
	\tablefoot{The methods are based on Equations~\ref{eq4}, \ref{eq5}, and \ref{eq6}, for M$_{\rm BH}$; and on Equations~\ref{eq8}, \ref{eq9}, and \ref{eq10}, for R$_{\rm Edd}$. In the last column the continuum luminosity used to calculate the Eddington ratio.}
\end{table}

\begin{table}
\caption{Geometric parameters for PMN J0948+0022.}
\centering
	\begin{tabular}{l|lll}
            \hline
		 {\bf Spectrum} & {\bf R$_{\rm{sub}}$} & {\bf R$_{\rm{out}}$} & {\bf R$_{\rm{NLR}}$} \\
		  & {\bf [pc]} & {\bf [pc]} & {\bf [pc]} \\
            \hline
            \hline
		SDSS		& 0.50 $\pm$ 0.02 & 15.1 $\pm$ 0.7 & $<$ 852 \\
		X-Shooter		& 0.51 $\pm$ 0.02 & 15.3 $\pm$ 0.7 & 485 $\pm$ 45 \\
		MUSE		& 0.50 $\pm$ 0.02 & 15.1 $\pm$ 0.7 & 742 $\pm$ 17	\\
	  \hline
	  \hline
	     	Reference 	& 0.50 $\pm$ 0.02 & 15.2 $\pm$ 0.7 & 556 $\pm$ 48\\
            \hline
	\end{tabular}
	\label{tab4}
	\tablefoot{The parameters were calculated using Equations \ref{eq11}, \ref{eq12}, and \ref{eq13}. R$_{\rm NLR}$ is estimated through L$_{\rm [O~III]}$, for SDSS the [O III] lines are upper limits, and consequently the derived parameter.}
\end{table}

\section{Discussion}\label{sec5}
\begin{figure}
    \centering
    \includegraphics[width=0.49\textwidth]{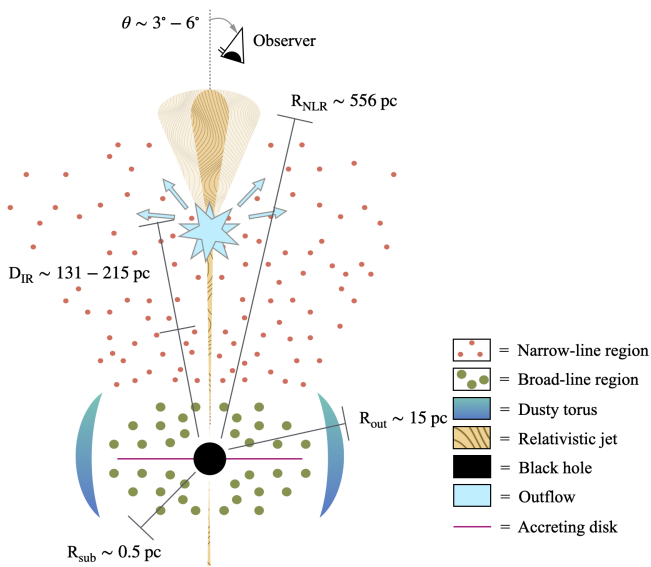} 
    \caption{Schematic representation of PMN J0948+0022. The scheme is not to scale.}
    \label{fig4}
\end{figure}

\begin{figure}
    \centering
    \includegraphics[width=0.45\textwidth]{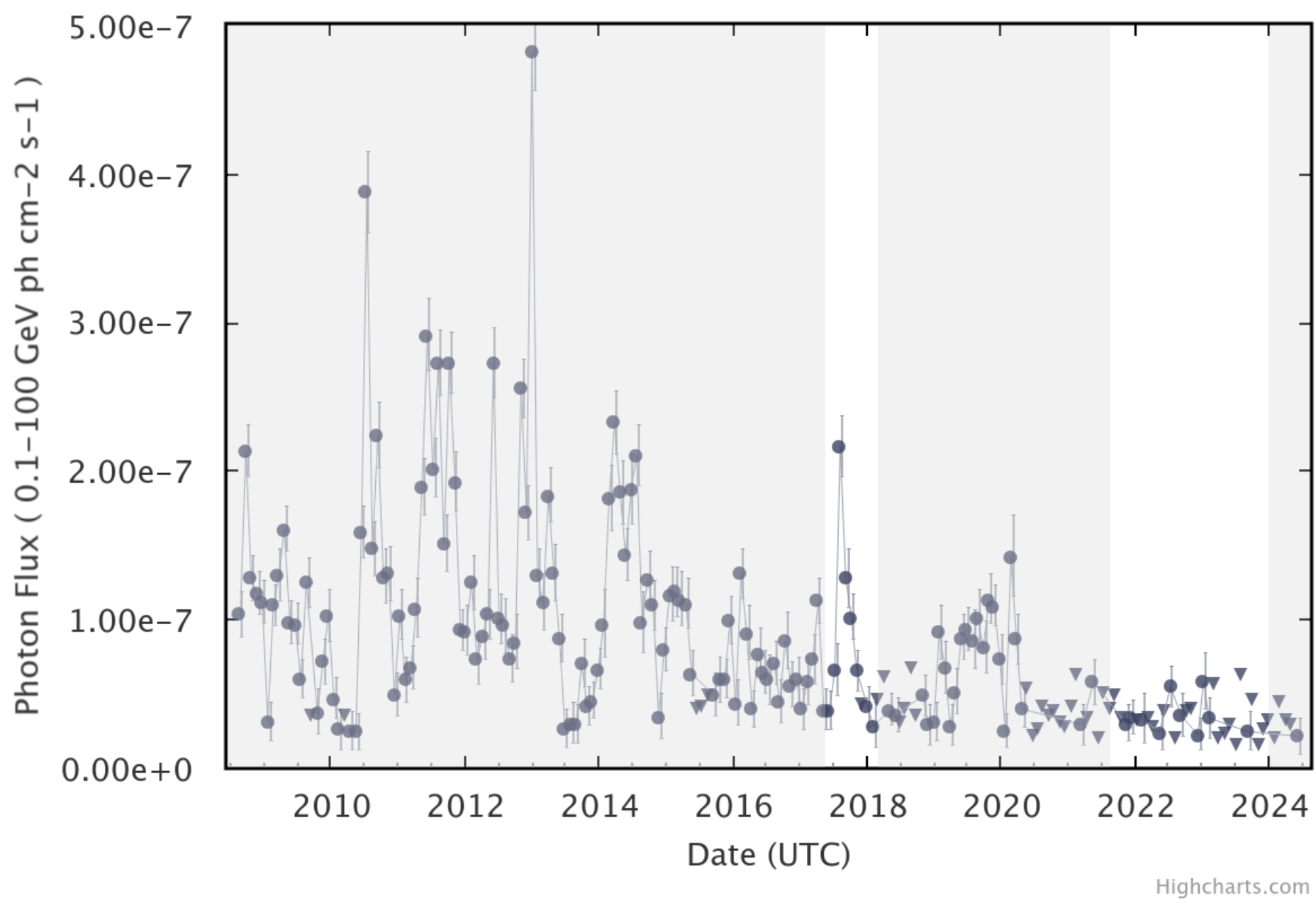} 
    \caption{Fermi light curve for PMN J0948+0022 with a data binning of 1 month. We highlighted the two periods that coincide with the X-Shooter and MUSE observations. The highlighted peak was observed on August 2, 2017, while the quiet phase began on April 18, 2020, and is still ongoing.}
    \label{fig5}
\end{figure}

Variations in optical line properties are generally attributed to changes in the accretion activity of the disk. However, this explanation only partially applies to the case of PMN J0948+0022. The R$_{\rm{Edd}}$ parameter, reported in Table~\ref{tab3} and calculated using Equation~\ref{eq9}, remains consistent across different observations. Spectral variations driven by accretion activity typically emerge first in the BLR (after $\sim46$ days, in our case) and propagate to the NLR over longer timescales ($\sim 2700$ years, for PMN J0948+0022). In our case, the flux of H$\beta_{\rm b}$ remains stable over time, although we observed changes in H$\beta_{\rm b1}$ and H$\beta_{\rm b2}$, which suggest an adjustment of the stratification. Conversely, we observed changes in the H$\beta_{\rm n}$ and [O III]$\lambda$5007. 

As noted in Section~\ref{sec3.3}, the better spectral resolution of X-Shooter and MUSE, requires a change in the H$\beta$ line profile from a Lorentzian to a combination of three Gaussians, which is typical of intermediate Seyferts. This type of reclassification is not unprecedented. For instance, SDSS J211852.96-073227.5 and SDSS J164100.10+345452.7 were initially classified as NLS1, and later reclassified as IS following new observations (\citeads{2020A&A...636L..12J};\citeads{2025A&A...696A..74C}). However, as we will show, PMN J0948+0022 is a different case. 

According to the UM, IS are viewed at large angles, and consequently the composite profile is due to partial obscuration. This is the case of SDSS J164100.10+345452.7 -- reclassified as IS by\citeads{2025A&A...696A..74C} -- which displays intrinsic obscuration in X$-$rays (\citeads{2023A&A...673A..85R}). In contrast, many authors (\citeads{2015A&A...575A..13F};\citeads{2014MNRAS.440..106B};\citeads{2014MNRAS.438.3521D};\citeads{2019A&A...632A.120B}) found that PMN J0948+0022 does not show any evidence of intrinsic obscuration in X$-$rays. 

The optical spectra analyzed in the present work also support the absence of intrinsic absorption. In particular, the detection of prominent iron multiplets ($R_{4570} \gtrsim 1$) -- thought to originate in the innermost regions of the NLR, close to the accretion disk and within the obscuring torus\citepads{2010ApJ...721L.143D} -- further supports this scenario. While the Fe II bump strength shows some dependence on inclination, this effect is considered secondary (\citeads{2023A&A...678A..63S};\citeads{2020CoSka..50..293P}) and should be interpreted as such. 

Another point refers to the effect of the obscuration on the FWHM of the broad lines. In this case, we would expect some difference between H$\beta$ at 4861\AA~and H$\alpha$ at 6563\AA, since the latter is in the red part of the spectrum. However, we found consistent values within 2$\sigma$: FWHM(H$\beta_{\rm b}$)=1520$\pm$31 km s$^{-1}$ and FWHM(H$\alpha_{\rm b}$)=1466$\pm$20 km s$^{-1}$.

In addition, we know the viewing angle of the relativistic jet to be about $3^{\circ}-6^{\circ}$ (\citeads{2009ApJ...707L.142A};\citeads{2012A&A...548A.106F};\citeads{2019MNRAS.487..640D}). Since it is known that there is a strong correlation between the jet and the NLR-bicone axes, while there is no correlation with the host-galaxy axis (\citeads{2003ApJ...597..768S};\citeads{2003ApJS..148..327S};\citeads{2013ApJS..209....1F};\citeads{2014ApJ...785...25F}), the small jet viewing angle confirms that we are observing the object almost face-on. There are cases of misalignment between the AGN and the host galaxy (\citeads{2019A&A...623A.172C};\citeads{2020A&A...639A...5M}), see also the case of NGC 5506 described in (\citeads{2002A&A...391L..21N};\citeads{2013ApJS..209....1F};\citeads{2014ApJ...785...25F}), but -- again -- in such a case we should observe an additional absorption at X$-$rays, which is not the case. Furthermore, such a misalignment would likely be a result of a recent merger, which is ruled out by\citeads{2020MNRAS.492.1450O}.

If no obscuration in PMN J0948+0022 is affecting the spectral lines, another mechanism must be considered to explain the observed variations. A starting point is to focus on the changes in the H$\beta_{\rm n}$ and [O III] lines. Specifically, the changes of H$\beta_{\rm n}$ on timescales of $\sim$10--27 years, much smaller than those expected from the NLR ($\sim$2700 years) point to a mechanism different from the AGN ionization. Given the presence of a relativistic jet, it is straightforward to study the interaction between the jet and the NLR (e.g.,\citeads{1991Natur.349..138R};\citeads{2013MNRAS.433..622M};\citeads{2016A&A...591A..88B};\citeads{2022MNRAS.516..766M};\citeads{2025MNRAS.536.1166E}).  

By studying the radio morphology and emission of PMN J0948+0022,\citeads{2019MNRAS.487..640D} found that the relativistic jet changed its shape, from parabolic to conic, in the NLR (distance from the center 100$-$430 pc). This process is due to deceleration as the jet is converting part of its kinetic energy into internal energy. The observed changes in the [O III]$\lambda$5007 line suggest that this internal energy is then dissipated in the NLR. To prove this hypothesis, we can calculate the location where this occurs, by using the timescales obtained in Section~\ref{sec4.2}. The relationship between the timescale and the dimension of the emitting region $r_{\rm{IR}}$ is given by:

\begin{equation}
    \frac{\tau}{1+z} > \frac{r_{\rm{IR}}}{c \cdot \delta} \quad \rightarrow \quad r_{\rm{IR}} < \frac{\tau \cdot c \cdot \delta}{1+z}
    \label{eq14}
\end{equation}

\noindent where $c$ is the speed of light, $z$ is the redshift, $\tau=4.8^{+1.1}_{-0.7}$ years, and $\delta$ is the Doppler factor. After four years of observations of PMN J0948+0022,\citeads{2012A&A...548A.106F} reported changes of the bulk Lorentz factor ($\Gamma$) between 11 and 16. By assuming a viewing angle of 3$^{\circ}$, it results $\delta$=16.5$-$18.8. Given the self-similarity of the jet and assuming a typical scaling factor of 0.1, then the distance from the central BH is:

\begin{equation}
    D_{\rm{IR}} = \frac{r_{\rm{IR}}}{0.1} \sim 131-215~\mathrm{pc}
    \label{eq15}
\end{equation}

\citeads{2021ApJ...923...67H} reported $\delta=47$ on the basis of Very Long Baseline Interferometer (VLBI) radio observations at 15 GHz. By assuming this value, and $\tau=4.8$ years, D$_{\rm IR}\sim435$ pc. All these values are consistent with the distance found by\citeads{2019MNRAS.487..640D}, thus strengthening our hypothesis that we are observing at optical wavelengths the transfer of the jet internal energy to the NLR. Since the [O III]$\lambda5007_{\rm c}$ increases from X-Shooter to MUSE observations, we expect a corresponding decrease of the jet activity. This is precisely what we observe at high-energy in the $\gamma-$ray (0.1-100 GeV) light curve from Fermi Large Area Telescope (LAT)\footnote{\url{https://fermi.gsfc.nasa.gov/ssc/data/access/lat/LightCurveRepository/source.html?source_name=4FGL_J0948.9+0022}}, displayed in Fig.~\ref{fig5}. The binning is one month, to enhance the long-term activity periods. Since high-energy $\gamma$-rays are a clear product of the relativistic jet, it is evident from this plot that, after a significant activity period, the source has been in a quiet state since 2020. Specifically, the activity showed an notable peak around 2017/2018. Comparing the periods in which the X-Shooter and MUSE observations were carried out, we note an almost order of magnitude difference in the $\gamma$ photon flux ($\sim 2 \cdot 10^{-7}$ ph cm$^{-2}$ s$^{-1}$ in August 2017 versus $\sim 6 \cdot 10^{-8}$ ph cm$^{-2}$ s$^{-1}$ in January 2023). For the SDSS period, we do not have any data points because the Fermi satellite was not operating at that time.

A similar, but opposite, effect is also visible in the [O III]$\lambda5007_{\rm o}$ component, which displays: a flux decrease, an increase in its velocity ($\Delta v\sim 200$ km s$^{-1}$) together with a strengthening of the asymmetry (see Fig.~\ref{fig3}). All these variations occurred in a timescale ($\sim10$ years) which is consistent with that of H$\beta_{\rm n}$ one ($\tau\sim 10-27$). 

To summarize, we propose the following scenario: according to radio observations\citepads{2019MNRAS.487..640D}, at a distance of about 100--430 pc from the central BH the jet decelerates, changes its shape from parabolic to conical, and converts part of its kinetic energy into internal energy. From the optical observations, presented in this work, we suggest that the jet internal energy is transferred to the NLR (increase of [O III]$\lambda5007_{\rm c}$ flux), which generates an outflow propagating along the NLR and affecting H$\beta_{\rm n}$, see Fig.~\ref{fig4}.

\section{Conclusions}\label{sec6}
We analyzed three optical spectra of PMN J0948+0022 taken at different epochs, specifically in 2000 (SDSS), 2017 (X-Shooter), and 2022/2023 (MUSE, composition). The source exhibits multiwavelength variability, but we focused on the H$\beta$ and [O III]$\lambda5007$ spectral lines. We tested the robustness of the variations in these lines by reducing the spectral resolution of the X-Shooter spectrum to match that of the SDSS and MUSE spectra. We found: 

\begin{itemize}
	\item Despite being initially classified as a NLS1 based on the SDSS spectrum\citepads{2003ApJ...584..147Z}, the higher resolution of X-Shooter and MUSE supports its reclassification as an intermediate Seyfert (see Section~\ref{sec3.3}); however, the H$\beta$ composite profile is not due to a geometrical effect of a partial obscuration, but to the interaction of the relativistic jet with the NLR. 
	\item Although the total flux of the broad component of H$\beta$ is stable during all the observations (variations within 1$\sigma$), there are changes in the two sub-components H$\beta_{\rm b1}$ and H$\beta_{\rm b2}$, which imply a stratification of the BLR.
	\item We observe significant variation in the H$\beta_{\rm n}$ (a $\sim4\sigma$ variation in flux between SDSS and X-Shooter, and of $\sim3\sigma$ between X-Shooter and MUSE) and [O III]$\lambda$5007 line flux and asymmetry (variation $>8\sigma$ in flux between X-Shooter and MUSE), which we interpret as the interaction of the relativistic jet with the NLR, confirming what has been found by\citeads{2019MNRAS.487..640D} based on radio observations. The relativistic jet decelerates in the NLR (131$-$215 pc from the central BH), changes its shape from parabolic to conic and converts part of its kinetic energy into internal energy. This internal energy is then dissipated in the NLR, generating an outflow revealed by the blue wing in the [O III] line. 
	\item Based on these new high-quality spectra and considering the lack of additional obscuration, we recalculated M$_{\rm BH}=10^{7.76\pm0.12}$ M$_{\odot}$, and  R$_{\rm Edd}=(0.21\pm0.06)$. 
\end{itemize}

Finally, we emphasize the importance of high-spectral-resolution optical spectroscopy, to be performed with instruments such as X-Shooter or its successor Son of X-Shooter (SOXS), in order to gain deeper insights into the jet-NLR interaction. This approach would allow for a better understanding of the possible connection between high-energy processes -- such as the $\gamma$-ray emission from AGN -- and their optical counterpart.

\begin{acknowledgements}
The authors thank Dr. N. Winkel for the fruitful discussion during the MUSE-10yrs workshop (Garching bei M\"unchen, November 18-22).\\
The authors thank the referee for the detailed and helpful review of the manuscript, which was significantly improved by the received comments. \\
G.V. acknowledges support from the European Union (ERC, WINGS,101040227).\\
Based on observations collected at the European Southern Observatory under ESO programmes 099.B-0785(A) and 110.23WR.001. \\
Funding for the Sloan Digital Sky Survey IV has been provided by the Alfred P. Sloan Foundation, the U.S. Department of Energy Office of Science, and the Participating Institutions. SDSS acknowledges support and resources from the Center for High-Performance Computing at the University of Utah. The SDSS web site is \url{www.sdss4.org}. SDSS is managed by the Astrophysical Research Consortium for the Participating Institutions of the SDSS Collaboration including the Brazilian Participation Group, the Carnegie Institution for Science, Carnegie Mellon University, Center for Astrophysics | Harvard \& Smithsonian (CfA), the Chilean Participation Group, the French Participation Group, Instituto de Astrofísica de Canarias, The Johns Hopkins University, Kavli Institute for the Physics and Mathematics of the Universe (IPMU) / University of Tokyo, the Korean Participation Group, Lawrence Berkeley National Laboratory, Leibniz Institut für Astrophysik Potsdam (AIP), Max-Planck-Institut für Astronomie (MPIA Heidelberg), Max-Planck-Institut für Astrophysik (MPA Garching), Max-Planck-Institut für Extraterrestrische Physik (MPE), National Astronomical Observatories of China, New Mexico State University, New York University, University of Notre Dame, Observatório Nacional / MCTI, The Ohio State University, Pennsylvania State University, Shanghai Astronomical Observatory, United Kingdom Participation Group, Universidad Nacional Autónoma de México, University of Arizona, University of Colorado Boulder, University of Oxford, University of Portsmouth, University of Utah, University of Virginia, University of Washington, University of Wisconsin, Vanderbilt University, and Yale University.\\
This research has made use of the NASA/IPAC Extragalactic Database (NED), which is operated by the Jet Propulsion Laboratory, California Institute of Technology, under contract with the National Aeronautics and Space Administration.
\end{acknowledgements}

\bibliographystyle{aa}
\bibliography{biblio}

\begin{thebibliography}{94}
\expandafter\ifx\csname natexlab\endcsname\relax\def\natexlab#1{#1}\fi

\bibitem[{{Abazajian} {et~al.}(2009){Abazajian}, {Adelman-McCarthy},
  {Ag{\"u}eros}, {Allam}, {Allende Prieto}, {An}, {Anderson}, {Anderson},
  {Annis}, {Bahcall}, {Bailer-Jones}, {Barentine}, {Bassett}, {Becker},
  {Beers}, {Bell}, {Belokurov}, {Berlind}, {Berman}, {Bernardi}, {Bickerton},
  {Bizyaev}, {Blakeslee}, {Blanton}, {Bochanski}, {Boroski}, {Brewington},
  {Brinchmann}, {Brinkmann}, {Brunner}, {Budav{\'a}ri}, {Carey}, {Carliles},
  {Carr}, {Castander}, {Cinabro}, {Connolly}, {Csabai}, {Cunha}, {Czarapata},
  {Davenport}, {de Haas}, {Dilday}, {Doi}, {Eisenstein}, {Evans}, {Evans},
  {Fan}, {Friedman}, {Frieman}, {Fukugita}, {G{\"a}nsicke}, {Gates},
  {Gillespie}, {Gilmore}, {Gonzalez}, {Gonzalez}, {Grebel}, {Gunn},
  {Gy{\"o}ry}, {Hall}, {Harding}, {Harris}, {Harvanek}, {Hawley}, {Hayes},
  {Heckman}, {Hendry}, {Hennessy}, {Hindsley}, {Hoblitt}, {Hogan}, {Hogg},
  {Holtzman}, {Hyde}, {Ichikawa}, {Ichikawa}, {Im}, {Ivezi{\'c}}, {Jester},
  {Jiang}, {Johnson}, {Jorgensen}, {Juri{\'c}}, {Kent}, {Kessler}, {Kleinman},
  {Knapp}, {Konishi}, {Kron}, {Krzesinski}, {Kuropatkin}, {Lampeitl},
  {Lebedeva}, {Lee}, {Lee}, {French Leger}, {L{\'e}pine}, {Li}, {Lima}, {Lin},
  {Long}, {Loomis}, {Loveday}, {Lupton}, {Magnier}, {Malanushenko},
  {Malanushenko}, {Mandelbaum}, {Margon}, {Marriner}, {Mart{\'\i}nez-Delgado},
  {Matsubara}, {McGehee}, {McKay}, {Meiksin}, {Morrison}, {Mullally}, {Munn},
  {Murphy}, {Nash}, {Nebot}, {Neilsen}, {Newberg}, {Newman}, {Nichol},
  {Nicinski}, {Nieto-Santisteban}, {Nitta}, {Okamura}, {Oravetz}, {Ostriker},
  {Owen}, {Padmanabhan}, {Pan}, {Park}, {Pauls}, {Peoples}, {Percival}, {Pier},
  {Pope}, {Pourbaix}, {Price}, {Purger}, {Quinn}, {Raddick}, {Re Fiorentin},
  {Richards}, {Richmond}, {Riess}, {Rix}, {Rockosi}, {Sako}, {Schlegel},
  {Schneider}, {Scholz}, {Schreiber}, {Schwope}, {Seljak}, {Sesar}, {Sheldon},
  {Shimasaku}, {Sibley}, {Simmons}, {Sivarani}, {Allyn Smith}, {Smith},
  {Smol{\v{c}}i{\'c}}, {Snedden}, {Stebbins}, {Steinmetz}, {Stoughton},
  {Strauss}, {SubbaRao}, {Suto}, {Szalay}, {Szapudi}, {Szkody}, {Tanaka},
  {Tegmark}, {Teodoro}, {Thakar}, {Tremonti}, {Tucker}, {Uomoto}, {Vanden
  Berk}, {Vandenberg}, {Vidrih}, {Vogeley}, {Voges}, {Vogt}, {Wadadekar},
  {Watters}, {Weinberg}, {West}, {White}, {Wilhite}, {Wonders}, {Yanny},
  {Yocum}, {York}, {Zehavi}, {Zibetti}, \& {Zucker}}]{2009ApJS..182..543A}
{Abazajian}, K.~N., {Adelman-McCarthy}, J.~K., {Ag{\"u}eros}, M.~A., {et~al.}
  2009, \apjs, 182, 543

\bibitem[{{Abdo} {et~al.}(2009{\natexlab{a}}){Abdo}, {Ackermann}, {Ajello},
  {Axelsson}, {Baldini}, {Ballet}, {Barbiellini}, {Bastieri}, {Baughman},
  {Bechtol}, {Bellazzini}, {Berenji}, {Bloom}, {Bonamente}, {Borgland},
  {Bregeon}, {Brez}, {Brigida}, {Bruel}, {Burnett}, {Caliandro}, {Cameron},
  {Caraveo}, {Casandjian}, {Cavazzuti}, {Cecchi}, {{\c{C}}elik}, {Celotti},
  {Chekhtman}, {Chiang}, {Ciprini}, {Claus}, {Cohen-Tanugi}, {Collmar},
  {Conrad}, {Costamante}, {Cutini}, {de Angelis}, {de Palma}, {Silva}, {Drell},
  {Dumora}, {Farnier}, {Favuzzi}, {Fegan}, {Focke}, {Fortin}, {Foschini},
  {Frailis}, {Fuhrmann}, {Fukazawa}, {Funk}, {Fusco}, {Gargano}, {Gehrels},
  {Germani}, {Giglietto}, {Giordano}, {Giroletti}, {Glanzman}, {Godfrey},
  {Grenier}, {Grove}, {Guillemot}, {Guiriec}, {Hanabata}, {Hays}, {Hughes},
  {Jackson}, {J{\'o}hannesson}, {Johnson}, {Johnson}, {Kadler}, {Kamae},
  {Katagiri}, {Kataoka}, {Kawai}, {Kerr}, {Kn{\"o}dlseder}, {Kocian}, {Kuss},
  {Lande}, {Latronico}, {Longo}, {Loparco}, {Lott}, {Lovellette}, {Lubrano},
  {Madejski}, {Makeev}, {Max-Moerbeck}, {Mazziotta}, {McConville}, {McEnery},
  {McGlynn}, {Meurer}, {Michelson}, {Mitthumsiri}, {Mizuno}, {Moiseev},
  {Monte}, {Monzani}, {Morselli}, {Moskalenko}, {Nestoras}, {Nolan}, {Norris},
  {Nuss}, {Ohsugi}, {Omodei}, {Orlando}, {Ormes}, {Paneque}, {Parent},
  {Pavlidou}, {Pelassa}, {Pepe}, {Pesce-Rollins}, {Piron}, {Porter},
  {Rain{\`o}}, {Rando}, {Razzano}, {Readhead}, {Reimer}, {Reposeur},
  {Richards}, {Rodriguez}, {Roth}, {Ryde}, {Sadrozinski}, {Sanchez}, {Sander},
  {Saz Parkinson}, {Scargle}, {Sgr{\`o}}, {Shaw}, {Smith}, {Spandre},
  {Spinelli}, {Strickman}, {Suson}, {Tagliaferri}, {Tajima}, {Takahashi},
  {Tanaka}, {Thayer}, {Thayer}, {Thompson}, {Tibaldo}, {Tibolla}, {Torres},
  {Tosti}, {Tramacere}, {Uchiyama}, {Usher}, {Vasileiou}, {Vilchez}, {Vitale},
  {Waite}, {Wang}, {Wehrle}, {Winer}, {Wood}, {Ylinen}, {Zensus}, {Ziegler},
  {Fermi/LAT Collaboration}, {Angelakis}, {Bailyn}, {Bignall}, {Blanchard},
  {Bonning}, {Buxton}, {Canterna}, {Carrami{\~n}ana}, {Carrasco}, {Colomer},
  {Doi}, {Ghisellini}, {Hauser}, {Hong}, {Isler}, {Kino}, {Kovalev}, {Kovalev},
  {Krichbaum}, {Kutyrev}, {Lahteenmaki}, {van Langevelde}, {Lister}, {Macomb},
  {Maraschi}, {Marchili}, {Nagai}, {Paragi}, {Phillips}, {Pushkarev},
  {Recillas}, {Roming}, {Sekido}, {Stark}, {Szomoru}, {Tammi}, {Tavecchio},
  {Tornikoski}, {Tzioumis}, {Urry}, \& {Wagner}}]{2009ApJ...707..727A}
{Abdo}, A.~A., {Ackermann}, M., {Ajello}, M., {et~al.} 2009{\natexlab{a}},
  \apj, 707, 727

\bibitem[{{Abdo} {et~al.}(2009{\natexlab{b}}){Abdo}, {Ackermann}, {Ajello},
  {Baldini}, {Ballet}, {Barbiellini}, {Bastieri}, {Bechtol}, {Bellazzini},
  {Berenji}, {Bloom}, {Bonamente}, {Borgland}, {Bregeon}, {Brez}, {Brigida},
  {Bruel}, {Burnett}, {Caliandro}, {Cameron}, {Caraveo}, {Casandjian},
  {Cecchi}, {{\c{C}}elik}, {Chekhtman}, {Cheung}, {Chiang}, {Ciprini}, {Claus},
  {Cohen-Tanugi}, {Conrad}, {Cutini}, {Dermer}, {de Palma}, {Silva}, {Drell},
  {Dubois}, {Dumora}, {Farnier}, {Favuzzi}, {Fegan}, {Focke}, {Foschini},
  {Frailis}, {Fukazawa}, {Fusco}, {Gargano}, {Gehrels}, {Germani}, {Giebels},
  {Giglietto}, {Giordano}, {Giroletti}, {Glanzman}, {Godfrey}, {Grenier},
  {Grove}, {Guillemot}, {Guiriec}, {Hayashida}, {Hays}, {Horan}, {Hughes},
  {J{\'o}hannesson}, {Johnson}, {Johnson}, {Kadler}, {Kamae}, {Katagiri},
  {Kataoka}, {Kerr}, {Kn{\"o}dlseder}, {Kuss}, {Lande}, {Latronico}, {Longo},
  {Loparco}, {Lott}, {Lovellette}, {Lubrano}, {Makeev}, {Mazziotta},
  {McConville}, {McEnery}, {Meurer}, {Michelson}, {Mitthumsiri}, {Mizuno},
  {Monte}, {Monzani}, {Morselli}, {Moskalenko}, {Murgia}, {Nolan}, {Norris},
  {Nuss}, {Ohsugi}, {Omodei}, {Orlando}, {Ormes}, {Pelassa}, {Pepe}, {Persic},
  {Pesce-Rollins}, {Piron}, {Porter}, {Rain{\`o}}, {Rando}, {Razzano},
  {Rochester}, {Rodriguez}, {Ryde}, {Sadrozinski}, {Sambruna}, {Sander}, {Saz
  Parkinson}, {Scargle}, {Sgr{\`o}}, {Smith}, {Spandre}, {Spinelli},
  {Strickman}, {Suson}, {Tagliaferri}, {Takahashi}, {Takahashi}, {Tanaka},
  {Thayer}, {Thayer}, {Thompson}, {Tibaldo}, {Tibolla}, {Torres}, {Tosti},
  {Tramacere}, {Uchiyama}, {Usher}, {Vasileiou}, {Vilchez}, {Vitale}, {Waite},
  {Wang}, {Winer}, {Wood}, {Ylinen}, {Ziegler}, {Fermi/LAT Collaboration},
  {Ghisellini}, {Maraschi}, \& {Tavecchio}}]{2009ApJ...707L.142A}
{Abdo}, A.~A., {Ackermann}, M., {Ajello}, M., {et~al.} 2009{\natexlab{b}},
  \apjl, 707, L142

\bibitem[{{Ahn} {et~al.}(2012){Ahn}, {Alexandroff}, {Allende Prieto},
  {Anderson}, {Anderton}, {Andrews}, {Aubourg}, {Bailey}, {Balbinot}, {Barnes},
  {Bautista}, {Beers}, {Beifiori}, {Berlind}, {Bhardwaj}, {Bizyaev}, {Blake},
  {Blanton}, {Blomqvist}, {Bochanski}, {Bolton}, {Borde}, {Bovy}, {Brandt},
  {Brinkmann}, {Brown}, {Brownstein}, {Bundy}, {Busca}, {Carithers}, {Carnero},
  {Carr}, {Casetti-Dinescu}, {Chen}, {Chiappini}, {Comparat}, {Connolly},
  {Crepp}, {Cristiani}, {Croft}, {Cuesta}, {da Costa}, {Davenport}, {Dawson},
  {de Putter}, {De Lee}, {Delubac}, {Dhital}, {Ealet}, {Ebelke}, {Edmondson},
  {Eisenstein}, {Escoffier}, {Esposito}, {Evans}, {Fan}, {Femen{\'\i}a
  Castell{\'a}}, {Fern{\'a}ndez Alvar}, {Ferreira}, {Filiz Ak}, {Finley},
  {Fleming}, {Font-Ribera}, {Frinchaboy}, {Garc{\'\i}a-Hern{\'a}ndez},
  {Garc{\'\i}a P{\'e}rez}, {Ge}, {G{\'e}nova-Santos}, {Gillespie}, {Girardi},
  {Gonz{\'a}lez Hern{\'a}ndez}, {Grebel}, {Gunn}, {Guo}, {Haggard}, {Hamilton},
  {Harris}, {Hawley}, {Hearty}, {Ho}, {Hogg}, {Holtzman}, {Honscheid},
  {Huehnerhoff}, {Ivans}, {Ivezi{\'c}}, {Jacobson}, {Jiang}, {Johansson},
  {Johnson}, {Kauffmann}, {Kirkby}, {Kirkpatrick}, {Klaene}, {Knapp}, {Kneib},
  {Le Goff}, {Leauthaud}, {Lee}, {Lee}, {Long}, {Loomis}, {Lucatello},
  {Lundgren}, {Lupton}, {Ma}, {Ma}, {MacDonald}, {Mack}, {Mahadevan}, {Maia},
  {Majewski}, {Makler}, {Malanushenko}, {Malanushenko}, {Manchado},
  {Mandelbaum}, {Manera}, {Maraston}, {Margala}, {Martell}, {McBride},
  {McGreer}, {McMahon}, {M{\'e}nard}, {Meszaros}, {Miralda-Escud{\'e}},
  {Montero-Dorta}, {Montesano}, {Morrison}, {Muna}, {Munn}, {Murayama},
  {Myers}, {Neto}, {Nguyen}, {Nichol}, {Nidever}, {Noterdaeme}, {Nuza},
  {Ogando}, {Olmstead}, {Oravetz}, {Owen}, {Padmanabhan},
  {Palanque-Delabrouille}, {Pan}, {Parejko}, {Parihar}, {P{\^a}ris},
  {Pattarakijwanich}, {Pepper}, {Percival}, {P{\'e}rez-Fournon},
  {P{\'e}rez-R{\`a}fols}, {Petitjean}, {Pforr}, {Pieri}, {Pinsonneault}, {Porto
  de Mello}, {Prada}, {Price-Whelan}, {Raddick}, {Rebolo}, {Rich}, {Richards},
  {Robin}, {Rocha-Pinto}, {Rockosi}, {Roe}, {Ross}, {Ross}, {Rossi},
  {Rubi{\~n}o-Martin}, {Samushia}, {Sanchez Almeida}, {S{\'a}nchez},
  {Santiago}, {Sayres}, {Schlegel}, {Schlesinger}, {Schmidt}, {Schneider},
  {Schultheis}, {Schwope}, {Sc{\'o}ccola}, {Seljak}, {Sheldon}, {Shen}, {Shu},
  {Simmerer}, {Simmons}, {Skibba}, {Skrutskie}, {Slosar}, {Sobreira}, {Sobeck},
  {Stassun}, {Steele}, {Steinmetz}, {Strauss}, {Streblyanska}, {Suzuki},
  {Swanson}, {Tal}, {Thakar}, {Thomas}, {Thompson}, {Tinker}, {Tojeiro},
  {Tremonti}, {Vargas Maga{\~n}a}, {Verde}, {Viel}, {Vikas}, {Vogt}, {Wake},
  {Wang}, {Weaver}, {Weinberg}, {Weiner}, {West}, {White}, {Wilson},
  {Wisniewski}, {Wood-Vasey}, {Yanny}, {Y{\`e}che}, {York}, {Zamora},
  {Zasowski}, {Zehavi}, {Zhao}, {Zheng}, {Zhu}, \&
  {Zinn}}]{2012ApJS..203...21A}
{Ahn}, C.~P., {Alexandroff}, R., {Allende Prieto}, C., {et~al.} 2012, \apjs,
  203, 21

\bibitem[{{Ahumada} {et~al.}(2020){Ahumada}, {Allende Prieto}, {Almeida},
  {Anders}, {Anderson}, {Andrews}, {Anguiano}, {Arcodia}, {Armengaud},
  {Aubert}, {Avila}, {Avila-Reese}, {Badenes}, {Balland}, {Barger},
  {Barrera-Ballesteros}, {Basu}, {Bautista}, {Beaton}, {Beers}, {Benavides},
  {Bender}, {Bernardi}, {Bershady}, {Beutler}, {Bidin}, {Bird}, {Bizyaev},
  {Blanc}, {Blanton}, {Boquien}, {Borissova}, {Bovy}, {Brandt}, {Brinkmann},
  {Brownstein}, {Bundy}, {Bureau}, {Burgasser}, {Burtin}, {Cano-D{\'\i}az},
  {Capasso}, {Cappellari}, {Carrera}, {Chabanier}, {Chaplin}, {Chapman},
  {Cherinka}, {Chiappini}, {Doohyun Choi}, {Chojnowski}, {Chung}, {Clerc},
  {Coffey}, {Comerford}, {Comparat}, {da Costa}, {Cousinou}, {Covey}, {Crane},
  {Cunha}, {Ilha}, {Dai}, {Damsted}, {Darling}, {Davidson}, {Davies}, {Dawson},
  {De}, {de la Macorra}, {De Lee}, {Queiroz}, {Deconto Machado}, {de la Torre},
  {Dell'Agli}, {du Mas des Bourboux}, {Diamond-Stanic}, {Dillon}, {Donor},
  {Drory}, {Duckworth}, {Dwelly}, {Ebelke}, {Eftekharzadeh}, {Davis Eigenbrot},
  {Elsworth}, {Eracleous}, {Erfanianfar}, {Escoffier}, {Fan}, {Farr},
  {Fern{\'a}ndez-Trincado}, {Feuillet}, {Finoguenov}, {Fofie},
  {Fraser-McKelvie}, {Frinchaboy}, {Fromenteau}, {Fu}, {Galbany}, {Garcia},
  {Garc{\'\i}a-Hern{\'a}ndez}, {Garma Oehmichen}, {Ge}, {Geimba Maia},
  {Geisler}, {Gelfand}, {Goddy}, {Gonzalez-Perez}, {Grabowski}, {Green},
  {Grier}, {Guo}, {Guy}, {Harding}, {Hasselquist}, {Hawken}, {Hayes}, {Hearty},
  {Hekker}, {Hogg}, {Holtzman}, {Horta}, {Hou}, {Hsieh}, {Huber}, {Hunt}, {Ider
  Chitham}, {Imig}, {Jaber}, {Jimenez Angel}, {Johnson}, {Jones},
  {J{\"o}nsson}, {Jullo}, {Kim}, {Kinemuchi}, {Kirkpatrick}, {Kite}, {Klaene},
  {Kneib}, {Kollmeier}, {Kong}, {Kounkel}, {Krishnarao}, {Lacerna}, {Lan},
  {Lane}, {Law}, {Le Goff}, {Leung}, {Lewis}, {Li}, {Lian}, {Lin}, {Long},
  {Longa-Pe{\~n}a}, {Lundgren}, {Lyke}, {Mackereth}, {MacLeod}, {Majewski},
  {Manchado}, {Maraston}, {Martini}, {Masseron}, {Masters}, {Mathur},
  {McDermid}, {Merloni}, {Merrifield}, {M{\'e}sz{\'a}ros}, {Miglio}, {Minniti},
  {Minsley}, {Miyaji}, {Mohammad}, {Mosser}, {Mueller}, {Muna},
  {Mu{\~n}oz-Guti{\'e}rrez}, {Myers}, {Nadathur}, {Nair}, {Nandra}, {Correa do
  Nascimento}, {Nevin}, {Newman}, {Nidever}, {Nitschelm}, {Noterdaeme},
  {O'Connell}, {Olmstead}, {Oravetz}, {Oravetz}, {Osorio}, {Pace}, {Padilla},
  {Palanque-Delabrouille}, {Palicio}, {Pan}, {Pan}, {Parker}, {Paviot},
  {Peirani}, {Ram{\'r}ez}, {Penny}, {Percival}, {Perez-Fournon},
  {P{\'e}rez-R{\`a}fols}, {Petitjean}, {Pieri}, {Pinsonneault}, {Poovelil},
  {Povick}, {Prakash}, {Price-Whelan}, {Raddick}, {Raichoor}, {Ray}, {Rembold},
  {Rezaie}, {Riffel}, {Riffel}, {Rix}, {Robin}, {Roman-Lopes},
  {Rom{\'a}n-Z{\'u}{\~n}iga}, {Rose}, {Ross}, {Rossi}, {Rowlands}, {Rubin},
  {Salvato}, {S{\'a}nchez}, {S{\'a}nchez-Menguiano}, {S{\'a}nchez-Gallego},
  {Sayres}, {Schaefer}, {Schiavon}, {Schimoia}, {Schlafly}, {Schlegel},
  {Schneider}, {Schultheis}, {Schwope}, {Seo}, {Serenelli}, {Shafieloo},
  {Shamsi}, {Shao}, {Shen}, {Shetrone}, {Shirley}, {Silva Aguirre}, {Simon},
  {Skrutskie}, {Slosar}, {Smethurst}, {Sobeck}, {Sodi}, {Souto}, {Stark},
  {Stassun}, {Steinmetz}, {Stello}, {Stermer}, {Storchi-Bergmann},
  {Streblyanska}, {Stringfellow}, {Stutz}, {Su{\'a}rez}, {Sun},
  {Taghizadeh-Popp}, {Talbot}, {Tayar}, {Thakar}, {Theriault}, {Thomas},
  {Thomas}, {Tinker}, {Tojeiro}, {Toledo}, {Tremonti}, {Troup}, {Tuttle},
  {Unda-Sanzana}, {Valentini}, {Vargas-Gonz{\'a}lez}, {Vargas-Maga{\~n}a},
  {V{\'a}zquez-Mata}, {Vivek}, {Wake}, {Wang}, {Weaver}, {Weijmans}, {Wild},
  {Wilson}, {Wilson}, {Wolthuis}, {Wood-Vasey}, {Yan}, {Yang}, {Y{\`e}che},
  {Zamora}, {Zarrouk}, {Zasowski}, {Zhang}, {Zhao}, {Zhao}, {Zheng}, {Zheng},
  {Zhu}, \& {Zou}}]{2020ApJS..249....3A}
{Ahumada}, R., {Allende Prieto}, C., {Almeida}, A., {et~al.} 2020, \apjs, 249,
  3

\bibitem[{{Alam} {et~al.}(2015){Alam}, {Albareti}, {Allende Prieto}, {Anders},
  {Anderson}, {Anderton}, {Andrews}, {Armengaud}, {Aubourg}, {Bailey}, {Basu},
  {Bautista}, {Beaton}, {Beers}, {Bender}, {Berlind}, {Beutler}, {Bhardwaj},
  {Bird}, {Bizyaev}, {Blake}, {Blanton}, {Blomqvist}, {Bochanski}, {Bolton},
  {Bovy}, {Shelden Bradley}, {Brandt}, {Brauer}, {Brinkmann}, {Brown},
  {Brownstein}, {Burden}, {Burtin}, {Busca}, {Cai}, {Capozzi}, {Carnero
  Rosell}, {Carr}, {Carrera}, {Chambers}, {Chaplin}, {Chen}, {Chiappini},
  {Chojnowski}, {Chuang}, {Clerc}, {Comparat}, {Covey}, {Croft}, {Cuesta},
  {Cunha}, {da Costa}, {Da Rio}, {Davenport}, {Dawson}, {De Lee}, {Delubac},
  {Deshpande}, {Dhital}, {Dutra-Ferreira}, {Dwelly}, {Ealet}, {Ebelke},
  {Edmondson}, {Eisenstein}, {Ellsworth}, {Elsworth}, {Epstein}, {Eracleous},
  {Escoffier}, {Esposito}, {Evans}, {Fan}, {Fern{\'a}ndez-Alvar}, {Feuillet},
  {Filiz Ak}, {Finley}, {Finoguenov}, {Flaherty}, {Fleming}, {Font-Ribera},
  {Foster}, {Frinchaboy}, {Galbraith-Frew}, {Garc{\'\i}a},
  {Garc{\'\i}a-Hern{\'a}ndez}, {Garc{\'\i}a P{\'e}rez}, {Gaulme}, {Ge},
  {G{\'e}nova-Santos}, {Georgakakis}, {Ghezzi}, {Gillespie}, {Girardi},
  {Goddard}, {Gontcho}, {Gonz{\'a}lez Hern{\'a}ndez}, {Grebel}, {Green},
  {Grieb}, {Grieves}, {Gunn}, {Guo}, {Harding}, {Hasselquist}, {Hawley},
  {Hayden}, {Hearty}, {Hekker}, {Ho}, {Hogg}, {Holley-Bockelmann}, {Holtzman},
  {Honscheid}, {Huber}, {Huehnerhoff}, {Ivans}, {Jiang}, {Johnson},
  {Kinemuchi}, {Kirkby}, {Kitaura}, {Klaene}, {Knapp}, {Kneib}, {Koenig},
  {Lam}, {Lan}, {Lang}, {Laurent}, {Le Goff}, {Leauthaud}, {Lee}, {Lee},
  {Licquia}, {Liu}, {Long}, {L{\'o}pez-Corredoira}, {Lorenzo-Oliveira},
  {Lucatello}, {Lundgren}, {Lupton}, {Mack}, {Mahadevan}, {Maia}, {Majewski},
  {Malanushenko}, {Malanushenko}, {Manchado}, {Manera}, {Mao}, {Maraston},
  {Marchwinski}, {Margala}, {Martell}, {Martig}, {Masters}, {Mathur},
  {McBride}, {McGehee}, {McGreer}, {McMahon}, {M{\'e}nard}, {Menzel},
  {Merloni}, {M{\'e}sz{\'a}ros}, {Miller}, {Miralda-Escud{\'e}}, {Miyatake},
  {Montero-Dorta}, {More}, {Morganson}, {Morice-Atkinson}, {Morrison},
  {Mosser}, {Muna}, {Myers}, {Nandra}, {Newman}, {Neyrinck}, {Nguyen},
  {Nichol}, {Nidever}, {Noterdaeme}, {Nuza}, {O'Connell}, {O'Connell},
  {O'Connell}, {Ogando}, {Olmstead}, {Oravetz}, {Oravetz}, {Osumi}, {Owen},
  {Padgett}, {Padmanabhan}, {Paegert}, {Palanque-Delabrouille}, {Pan},
  {Parejko}, {P{\^a}ris}, {Park}, {Pattarakijwanich}, {Pellejero-Ibanez},
  {Pepper}, {Percival}, {P{\'e}rez-Fournon}, {P{\'e}rez-R{\`a}fols},
  {Petitjean}, {Pieri}, {Pinsonneault}, {Porto de Mello}, {Prada}, {Prakash},
  {Price-Whelan}, {Protopapas}, {Raddick}, {Rahman}, {Reid}, {Rich}, {Rix},
  {Robin}, {Rockosi}, {Rodrigues}, {Rodr{\'\i}guez-Torres}, {Roe}, {Ross},
  {Ross}, {Rossi}, {Ruan}, {Rubi{\~n}o-Mart{\'\i}n}, {Rykoff},
  {Salazar-Albornoz}, {Salvato}, {Samushia}, {S{\'a}nchez}, {Santiago},
  {Sayres}, {Schiavon}, {Schlegel}, {Schmidt}, {Schneider}, {Schultheis},
  {Schwope}, {Sc{\'o}ccola}, {Scott}, {Sellgren}, {Seo}, {Serenelli}, {Shane},
  {Shen}, {Shetrone}, {Shu}, {Silva Aguirre}, {Sivarani}, {Skrutskie},
  {Slosar}, {Smith}, {Sobreira}, {Souto}, {Stassun}, {Steinmetz}, {Stello},
  {Strauss}, {Streblyanska}, {Suzuki}, {Swanson}, {Tan}, {Tayar}, {Terrien},
  {Thakar}, {Thomas}, {Thomas}, {Thompson}, {Tinker}, {Tojeiro}, {Troup},
  {Vargas-Maga{\~n}a}, {Vazquez}, {Verde}, {Viel}, {Vogt}, {Wake}, {Wang},
  {Weaver}, {Weinberg}, {Weiner}, {White}, {Wilson}, {Wisniewski},
  {Wood-Vasey}, {Ye`che}, {York}, {Zakamska}, {Zamora}, {Zasowski}, {Zehavi},
  {Zhao}, {Zheng}, {Zhou}, {Zhou}, {Zou}, \& {Zhu}}]{2015ApJS..219...12A}
{Alam}, S., {Albareti}, F.~D., {Allende Prieto}, C., {et~al.} 2015, \apjs, 219,
  12

\bibitem[{{Angelakis} {et~al.}(2015){Angelakis}, {Fuhrmann}, {Marchili},
  {Foschini}, {Myserlis}, {Karamanavis}, {Komossa}, {Blinov}, {Krichbaum},
  {Sievers}, {Ungerechts}, \& {Zensus}}]{2015A&A...575A..55A}
{Angelakis}, E., {Fuhrmann}, L., {Marchili}, N., {et~al.} 2015, \aap, 575, A55

\bibitem[{{Antonucci}(1993)}]{1993ARA&A..31..473A}
{Antonucci}, R. 1993, \araa, 31, 473

\bibitem[{{Baldwin} {et~al.}(1981){Baldwin}, {Phillips}, \&
  {Terlevich}}]{1981PASP...93....5B}
{Baldwin}, J.~A., {Phillips}, M.~M., \& {Terlevich}, R. 1981, \pasp, 93, 5

\bibitem[{{Bentz} {et~al.}(2013){Bentz}, {Denney}, {Grier}, {Barth},
  {Peterson}, {Vestergaard}, {Bennert}, {Canalizo}, {De Rosa}, {Filippenko},
  {Gates}, {Greene}, {Li}, {Malkan}, {Pogge}, {Stern}, {Treu}, \&
  {Woo}}]{2013ApJ...767..149B}
{Bentz}, M.~C., {Denney}, K.~D., {Grier}, C.~J., {et~al.} 2013, \apj, 767, 149

\bibitem[{{Berton} {et~al.}(2019){Berton}, {Braito}, {Mathur}, {Foschini},
  {Piconcelli}, {Chen}, \& {Pogge}}]{2019A&A...632A.120B}
{Berton}, M., {Braito}, V., {Mathur}, S., {et~al.} 2019, \aap, 632, A120

\bibitem[{{Berton} {et~al.}(2017){Berton}, {Foschini}, {Caccianiga}, {Ciroi},
  {Congiu}, {Cracco}, {Frezzato}, {La Mura}, \&
  {Rafanelli}}]{2017FrASS...4....8B}
{Berton}, M., {Foschini}, L., {Caccianiga}, A., {et~al.} 2017, Frontiers in
  Astronomy and Space Sciences, 4, 8

\bibitem[{{Berton} {et~al.}(2016){Berton}, {Foschini}, {Ciroi}, {Cracco}, {La
  Mura}, {Di Mille}, \& {Rafanelli}}]{2016A&A...591A..88B}
{Berton}, M., {Foschini}, L., {Ciroi}, S., {et~al.} 2016, \aap, 591, A88

\bibitem[{{Berton} {et~al.}(2021){Berton}, {Peluso}, {Marziani}, {Komossa},
  {Foschini}, {Ciroi}, {Chen}, {Congiu}, {Gallo}, {Bj{\"o}rklund}, {Crepaldi},
  {Di Mille}, {J{\"a}rvel{\"a}}, {Kotilainen}, {Kreikenbohm}, {Morrell},
  {Romano}, {Sani}, {Terreran}, {Tornikoski}, {Vercellone}, \&
  {Vietri}}]{2021A&A...654A.125B}
{Berton}, M., {Peluso}, G., {Marziani}, P., {et~al.} 2021, \aap, 654, A125

\bibitem[{{Bhattacharyya} {et~al.}(2014){Bhattacharyya}, {Bhatt}, {Bhatt}, \&
  {Singh}}]{2014MNRAS.440..106B}
{Bhattacharyya}, S., {Bhatt}, H., {Bhatt}, N., \& {Singh}, K.~K. 2014, \mnras,
  440, 106

\bibitem[{{Blandford} {et~al.}(2019){Blandford}, {Meier}, \&
  {Readhead}}]{2019ARA&A..57..467B}
{Blandford}, R., {Meier}, D., \& {Readhead}, A. 2019, \araa, 57, 467

\bibitem[{{Circosta} {et~al.}(2019){Circosta}, {Vignali}, {Gilli}, {Feltre},
  {Vito}, {Calura}, {Mainieri}, {Massardi}, \& {Norman}}]{2019A&A...623A.172C}
{Circosta}, C., {Vignali}, C., {Gilli}, R., {et~al.} 2019, \aap, 623, A172

\bibitem[{{Collin} {et~al.}(2006){Collin}, {Kawaguchi}, {Peterson}, \&
  {Vestergaard}}]{2006A&A...456...75C}
{Collin}, S., {Kawaguchi}, T., {Peterson}, B.~M., \& {Vestergaard}, M. 2006,
  \aap, 456, 75

\bibitem[{{Comparat} {et~al.}(2013){Comparat}, {Kneib}, {Bacon}, {Mostek},
  {Newman}, {Schlegel}, \& {Y{\`e}che}}]{2013A&A...559A..18C}
{Comparat}, J., {Kneib}, J.-P., {Bacon}, R., {et~al.} 2013, \aap, 559, A18

\bibitem[{{Cracco} {et~al.}(2016){Cracco}, {Ciroi}, {Berton}, {Di Mille},
  {Foschini}, {La Mura}, \& {Rafanelli}}]{2016MNRAS.462.1256C}
{Cracco}, V., {Ciroi}, S., {Berton}, M., {et~al.} 2016, \mnras, 462, 1256

\bibitem[{{Crepaldi} {et~al.}(2025){Crepaldi}, {Berton}, {Dalla Barba}, {La
  Mura}, {J{\"a}rvel{\"a}}, {Vietri}, \& {Ciroi}}]{2025A&A...696A..74C}
{Crepaldi}, L., {Berton}, M., {Dalla Barba}, B., {et~al.} 2025, \aap, 696, A74

\bibitem[{{D'Ammando} {et~al.}(2014){D'Ammando}, {Larsson}, {Orienti},
  {Raiteri}, {Angelakis}, {Carrami{\~n}ana}, {Carrasco}, {Drake}, {Fuhrmann},
  {Giroletti}, {Hovatta}, {Max-Moerbeck}, {Porras}, {Readhead}, {Recillas}, \&
  {Richards}}]{2014MNRAS.438.3521D}
{D'Ammando}, F., {Larsson}, J., {Orienti}, M., {et~al.} 2014, \mnras, 438, 3521

\bibitem[{{D'Ammando} {et~al.}(2015){D'Ammando}, {Orienti}, {Finke}, {Raiteri},
  {Hovatta}, {Larsson}, {Max-Moerbeck}, {Perkins}, {Readhead}, {Richards},
  {Beilicke}, {Benbow}, {Berger}, {Bird}, {Bugaev}, {Cardenzana}, {Cerruti},
  {Chen}, {Ciupik}, {Dickinson}, {Eisch}, {Errando}, {Falcone}, {Finley},
  {Fleischhack}, {Fortin}, {Fortson}, {Furniss}, {Gerard}, {Gillanders},
  {Griffiths}, {Grube}, {Gyuk}, {H{\r{a}}kansson}, {Holder}, {Humensky}, {Kar},
  {Kertzman}, {Khassen}, {Kieda}, {Krennrich}, {Kumar}, {Lang}, {Maier},
  {McCann}, {Meagher}, {Moriarty}, {Mukherjee}, {Nieto}, {de Bhr{\'o}ithe},
  {Ong}, {Otte}, {Pohl}, {Popkow}, {Prokoph}, {Pueschel}, {Quinn}, {Ragan},
  {Reynolds}, {Richards}, {Roache}, {Rousselle}, {Santander}, {Sembroski},
  {Smith}, {Staszak}, {Telezhinsky}, {Tucci}, {Tyler}, {Varlotta}, {Vassiliev},
  {Wakely}, {Weinstein}, {Welsing}, {Williams}, \&
  {Zitzer}}]{2015MNRAS.446.2456D}
{D'Ammando}, F., {Orienti}, M., {Finke}, J., {et~al.} 2015, \mnras, 446, 2456

\bibitem[{{Dimitrijevi{\'c}} {et~al.}(2007){Dimitrijevi{\'c}},
  {Kova{\v{c}}evi{\'c}}, {Popovi{\'c}}, {Da{\v{c}}i{\'c}}, \&
  {Ili{\'c}}}]{2007AIPC..895..313D}
{Dimitrijevi{\'c}}, M.~S., {Kova{\v{c}}evi{\'c}}, J., {Popovi{\'c}},
  L.~{\v{C}}., {Da{\v{c}}i{\'c}}, M., \& {Ili{\'c}}, D. 2007, in American
  Institute of Physics Conference Series, Vol. 895, Fifty Years of Romanian
  Astrophysics, ed. C.~{Dumitrache}, N.~A. {Popescu}, M.~D. {Suran}, \&
  V.~{Mioc} (AIP), 313--316

\bibitem[{{Doi} {et~al.}(2006){Doi}, {Nagai}, {Asada}, {Kameno}, {Wajima}, \&
  {Inoue}}]{2006PASJ...58..829D}
{Doi}, A., {Nagai}, H., {Asada}, K., {et~al.} 2006, \pasj, 58, 829

\bibitem[{{Doi} {et~al.}(2012){Doi}, {Nagira}, {Kawakatu}, {Kino}, {Nagai}, \&
  {Asada}}]{2012ApJ...760...41D}
{Doi}, A., {Nagira}, H., {Kawakatu}, N., {et~al.} 2012, \apj, 760, 41

\bibitem[{{Doi} {et~al.}(2019){Doi}, {Nakahara}, {Nakamura}, {Kino},
  {Kawakatu}, \& {Nagai}}]{2019MNRAS.487..640D}
{Doi}, A., {Nakahara}, S., {Nakamura}, M., {et~al.} 2019, \mnras, 487, 640

\bibitem[{{Doj{\v{c}}inovi{\'c}} {et~al.}(2023){Doj{\v{c}}inovi{\'c}},
  {Kova{\v{c}}evi{\'c}-Doj{\v{c}}inovi{\'c}}, \&
  {Popovi{\'c}}}]{2023AdSpR..71.1219D}
{Doj{\v{c}}inovi{\'c}}, I., {Kova{\v{c}}evi{\'c}-Doj{\v{c}}inovi{\'c}}, J., \&
  {Popovi{\'c}}, L.~{\v{C}}. 2023, Advances in Space Research, 71, 1219

\bibitem[{{Donato}(2010)}]{2010ATel.2733....1D}
{Donato}, D. 2010, The Astronomer's Telegram, 2733, 1

\bibitem[{{Dong} {et~al.}(2010){Dong}, {Ho}, {Wang}, {Wang}, {Wang}, {Fan}, \&
  {Zhou}}]{2010ApJ...721L.143D}
{Dong}, X.-B., {Ho}, L.~C., {Wang}, J.-G., {et~al.} 2010, \apjl, 721, L143

\bibitem[{{Drinkwater} {et~al.}(2018){Drinkwater}, {Byrne}, {Blake},
  {Glazebrook}, {Brough}, {Colless}, {Couch}, {Croton}, {Croom}, {Davis},
  {Forster}, {Gilbank}, {Hinton}, {Jelliffe}, {Jurek}, {Li}, {Martin},
  {Pimbblet}, {Poole}, {Pracy}, {Sharp}, {Smillie}, {Spolaor}, {Wisnioski},
  {Woods}, {Wyder}, \& {Yee}}]{2018MNRAS.474.4151D}
{Drinkwater}, M.~J., {Byrne}, Z.~J., {Blake}, C., {et~al.} 2018, \mnras, 474,
  4151

\bibitem[{{Du} \& {Wang}(2019)}]{2019ApJ...886...42D}
{Du}, P. \& {Wang}, J.-M. 2019, \apj, 886, 42

\bibitem[{{Duncan}(2022)}]{2022MNRAS.512.3662D}
{Duncan}, K.~J. 2022, \mnras, 512, 3662

\bibitem[{{Elitzur}(2008)}]{2008NewAR..52..274E}
{Elitzur}, M. 2008, \nar, 52, 274

\bibitem[{{Escott} {et~al.}(2025){Escott}, {Morabito}, {Scholtz}, {Hickox},
  {Harrison}, {Alexander}, {Arnaudova}, {Smith}, {Duncan}, {Petley},
  {Kondapally}, {Calistro Rivera}, \& {Kolwa}}]{2025MNRAS.536.1166E}
{Escott}, E.~L., {Morabito}, L.~K., {Scholtz}, J., {et~al.} 2025, \mnras, 536,
  1166

\bibitem[{{Fischer} {et~al.}(2013){Fischer}, {Crenshaw}, {Kraemer}, \&
  {Schmitt}}]{2013ApJS..209....1F}
{Fischer}, T.~C., {Crenshaw}, D.~M., {Kraemer}, S.~B., \& {Schmitt}, H.~R.
  2013, \apjs, 209, 1

\bibitem[{{Fischer} {et~al.}(2014){Fischer}, {Crenshaw}, {Kraemer}, {Schmitt},
  \& {Turner}}]{2014ApJ...785...25F}
{Fischer}, T.~C., {Crenshaw}, D.~M., {Kraemer}, S.~B., {Schmitt}, H.~R., \&
  {Turner}, T.~J. 2014, \apj, 785, 25

\bibitem[{{Fischer} {et~al.}(2018){Fischer}, {Kraemer}, {Schmitt}, {Longo
  Micchi}, {Crenshaw}, {Revalski}, {Vestergaard}, {Elvis}, {Gaskell}, {Hamann},
  {Ho}, {Hutchings}, {Mushotzky}, {Netzer}, {Storchi-Bergmann}, {Straughn},
  {Turner}, \& {Ward}}]{2018ApJ...856..102F}
{Fischer}, T.~C., {Kraemer}, S.~B., {Schmitt}, H.~R., {et~al.} 2018, \apj, 856,
  102

\bibitem[{{Foschini}(2020)}]{2020Univ....6..136F}
{Foschini}, L. 2020, Universe, 6, 136

\bibitem[{{Foschini} {et~al.}(2012){Foschini}, {Angelakis}, {Fuhrmann},
  {Ghisellini}, {Hovatta}, {Lahteenmaki}, {Lister}, {Braito}, {Gallo},
  {Hamilton}, {Kino}, {Komossa}, {Pushkarev}, {Thompson}, {Tibolla},
  {Tramacere}, {Carrami{\~n}ana}, {Carrasco}, {Falcone}, {Giroletti}, {Grupe},
  {Kovalev}, {Krichbaum}, {Max-Moerbeck}, {Nestoras}, {Pearson}, {Porras},
  {Readhead}, {Recillas}, {Richards}, {Riquelme}, {Sievers}, {Tammi},
  {Tornikoski}, {Ungerechts}, {Zensus}, {Celotti}, {Bonnoli}, {Doi},
  {Maraschi}, {Tagliaferri}, \& {Tavecchio}}]{2012A&A...548A.106F}
{Foschini}, L., {Angelakis}, E., {Fuhrmann}, L., {et~al.} 2012, \aap, 548, A106

\bibitem[{{Foschini} {et~al.}(2015){Foschini}, {Berton}, {Caccianiga}, {Ciroi},
  {Cracco}, {Peterson}, {Angelakis}, {Braito}, {Fuhrmann}, {Gallo}, {Grupe},
  {J{\"a}rvel{\"a}}, {Kaufmann}, {Komossa}, {Kovalev}, {L{\"a}hteenm{\"a}ki},
  {Lisakov}, {Lister}, {Mathur}, {Richards}, {Romano}, {Sievers},
  {Tagliaferri}, {Tammi}, {Tibolla}, {Tornikoski}, {Vercellone}, {La Mura},
  {Maraschi}, \& {Rafanelli}}]{2015A&A...575A..13F}
{Foschini}, L., {Berton}, M., {Caccianiga}, A., {et~al.} 2015, \aap, 575, A13

\bibitem[{{Foschini} {et~al.}(2010){Foschini}, {Fermi/Lat Collaboration},
  {Ghisellini}, {Maraschi}, {Tavecchio}, \& {Angelakis}}]{2010ASPC..427..243F}
{Foschini}, L., {Fermi/Lat Collaboration}, {Ghisellini}, G., {et~al.} 2010, in
  Astronomical Society of the Pacific Conference Series, Vol. 427, Accretion
  and Ejection in AGN: a Global View, ed. L.~{Maraschi}, G.~{Ghisellini},
  R.~{Della Ceca}, \& F.~{Tavecchio}, 243--248

\bibitem[{{Foschini} {et~al.}(2011){Foschini}, {Ghisellini}, {Kovalev},
  {Lister}, {D'Ammando}, {Thompson}, {Tramacere}, {Angelakis}, {Donato},
  {Falcone}, {Fuhrmann}, {Hauser}, {Kovalev}, {Mannheim}, {Maraschi},
  {Max-Moerbeck}, {Nestoras}, {Pavlidou}, {Pearson}, {Pushkarev}, {Readhead},
  {Richards}, {Stevenson}, {Tagliaferri}, {Tibolla}, {Tavecchio}, \&
  {Wagner}}]{2011MNRAS.413.1671F}
{Foschini}, L., {Ghisellini}, G., {Kovalev}, Y.~Y., {et~al.} 2011, \mnras, 413,
  1671

\bibitem[{{Foschini} {et~al.}(2022){Foschini}, {Lister}, {Andernach}, {Ciroi},
  {Marziani}, {Ant{\'o}n}, {Berton}, {Dalla Bont{\`a}}, {J{\"a}rvel{\"a}},
  {March{\~a}}, {Romano}, {Tornikoski}, {Vercellone}, \&
  {Vietri}}]{2022Univ....8..587F}
{Foschini}, L., {Lister}, M.~L., {Andernach}, H., {et~al.} 2022, Universe, 8,
  587

\bibitem[{{Fossati} {et~al.}(1998){Fossati}, {Maraschi}, {Celotti}, {Comastri},
  \& {Ghisellini}}]{1998MNRAS.299..433F}
{Fossati}, G., {Maraschi}, L., {Celotti}, A., {Comastri}, A., \& {Ghisellini},
  G. 1998, \mnras, 299, 433

\bibitem[{{Ghisellini} {et~al.}(1998){Ghisellini}, {Celotti}, {Fossati},
  {Maraschi}, \& {Comastri}}]{1998MNRAS.301..451G}
{Ghisellini}, G., {Celotti}, A., {Fossati}, G., {Maraschi}, L., \& {Comastri},
  A. 1998, \mnras, 301, 451

\bibitem[{{Giroletti} {et~al.}(2011){Giroletti}, {Paragi}, {Bignall}, {Doi},
  {Foschini}, {Gab{\'a}nyi}, {Reynolds}, {Blanchard}, {Campbell}, {Colomer},
  {Hong}, {Kadler}, {Kino}, {van Langevelde}, {Nagai}, {Phillips}, {Sekido},
  {Szomoru}, \& {Tzioumis}}]{2011A&A...528L..11G}
{Giroletti}, M., {Paragi}, Z., {Bignall}, H., {et~al.} 2011, \aap, 528, L11

\bibitem[{{Goodrich}(1989)}]{1989ApJ...342..224G}
{Goodrich}, R.~W. 1989, \apj, 342, 224

\bibitem[{{Greene} {et~al.}(2010){Greene}, {Hood}, {Barth}, {Bennert}, {Bentz},
  {Filippenko}, {Gates}, {Malkan}, {Treu}, {Walsh}, \&
  {Woo}}]{2010ApJ...723..409G}
{Greene}, J.~E., {Hood}, C.~E., {Barth}, A.~J., {et~al.} 2010, \apj, 723, 409

\bibitem[{{Homan} {et~al.}(2021){Homan}, {Cohen}, {Hovatta}, {Kellermann},
  {Kovalev}, {Lister}, {Popkov}, {Pushkarev}, {Ros}, \&
  {Savolainen}}]{2021ApJ...923...67H}
{Homan}, D.~C., {Cohen}, M.~H., {Hovatta}, T., {et~al.} 2021, \apj, 923, 67

\bibitem[{{Ili{\'c}} {et~al.}(2020){Ili{\'c}}, {Oknyansky}, {Popovi{\'c}},
  {Tsygankov}, {Belinski}, {Tatarnikov}, {Dodin}, {Shatsky}, {Ikonnikova},
  {Raki{\'c}}, {Kova{\v{c}}evi{\'c}}, {Mar{\v{c}}eta-Mandi{\'c}}, {Burlak},
  {Mishin}, {Metlova}, {Potanin}, \& {Zheltoukhov}}]{2020A&A...638A..13I}
{Ili{\'c}}, D., {Oknyansky}, V., {Popovi{\'c}}, L.~{\v{C}}., {et~al.} 2020,
  \aap, 638, A13

\bibitem[{{Ili{\'c}} {et~al.}(2023){Ili{\'c}}, {Raki{\'c}}, \&
  {Popovi{\'c}}}]{2023ApJS..267...19I}
{Ili{\'c}}, D., {Raki{\'c}}, N., \& {Popovi{\'c}}, L.~{\v{C}}. 2023, \apjs,
  267, 19

\bibitem[{{J{\"a}rvel{\"a}} {et~al.}(2020){J{\"a}rvel{\"a}}, {Berton}, {Ciroi},
  {Congiu}, {L{\"a}hteenm{\"a}ki}, \& {Di Mille}}]{2020A&A...636L..12J}
{J{\"a}rvel{\"a}}, E., {Berton}, M., {Ciroi}, S., {et~al.} 2020, \aap, 636, L12

\bibitem[{{Kaspi} {et~al.}(2000){Kaspi}, {Smith}, {Netzer}, {Maoz}, {Jannuzi},
  \& {Giveon}}]{2000ApJ...533..631K}
{Kaspi}, S., {Smith}, P.~S., {Netzer}, H., {et~al.} 2000, \apj, 533, 631

\bibitem[{{Kauffmann} {et~al.}(2003){Kauffmann}, {Heckman}, {Tremonti},
  {Brinchmann}, {Charlot}, {White}, {Ridgway}, {Brinkmann}, {Fukugita}, {Hall},
  {Ivezi{\'c}}, {Richards}, \& {Schneider}}]{2003MNRAS.346.1055K}
{Kauffmann}, G., {Heckman}, T.~M., {Tremonti}, C., {et~al.} 2003, \mnras, 346,
  1055

\bibitem[{{Keel}(1980)}]{1980AJ.....85..198K}
{Keel}, W.~C. 1980, \aj, 85, 198

\bibitem[{{Kewley} {et~al.}(2006){Kewley}, {Groves}, {Kauffmann}, \&
  {Heckman}}]{2006MNRAS.372..961K}
{Kewley}, L.~J., {Groves}, B., {Kauffmann}, G., \& {Heckman}, T. 2006, \mnras,
  372, 961

\bibitem[{{Komossa} {et~al.}(2008){Komossa}, {Xu}, {Zhou}, {Storchi-Bergmann},
  \& {Binette}}]{2008ApJ...680..926K}
{Komossa}, S., {Xu}, D., {Zhou}, H., {Storchi-Bergmann}, T., \& {Binette}, L.
  2008, \apj, 680, 926

\bibitem[{{L{\"a}hteenm{\"a}ki} {et~al.}(2017){L{\"a}hteenm{\"a}ki},
  {J{\"a}rvel{\"a}}, {Hovatta}, {Tornikoski}, {Harrison}, {L{\'o}pez-Caniego},
  {Max-Moerbeck}, {Mingaliev}, {Pearson}, {Ramakrishnan}, {Readhead}, {Reeves},
  {Richards}, {Sotnikova}, \& {Tammi}}]{2017A&A...603A.100L}
{L{\"a}hteenm{\"a}ki}, A., {J{\"a}rvel{\"a}}, E., {Hovatta}, T., {et~al.} 2017,
  \aap, 603, A100

\bibitem[{{Lister} {et~al.}(2016){Lister}, {Aller}, {Aller}, {Homan},
  {Kellermann}, {Kovalev}, {Pushkarev}, {Richards}, {Ros}, \&
  {Savolainen}}]{2016AJ....152...12L}
{Lister}, M.~L., {Aller}, M.~F., {Aller}, H.~D., {et~al.} 2016, \aj, 152, 12

\bibitem[{{Liu} {et~al.}(2010){Liu}, {Wang}, {Mao}, \&
  {Wei}}]{2010ApJ...715L.113L}
{Liu}, H., {Wang}, J., {Mao}, Y., \& {Wei}, J. 2010, \apjl, 715, L113

\bibitem[{{Malizia} {et~al.}(2020){Malizia}, {Bassani}, {Stephen}, {Bazzano},
  \& {Ubertini}}]{2020A&A...639A...5M}
{Malizia}, A., {Bassani}, L., {Stephen}, J.~B., {Bazzano}, A., \& {Ubertini},
  P. 2020, \aap, 639, A5

\bibitem[{{Mao}(2021)}]{2021RNAAS...5..109M}
{Mao}, L. 2021, Research Notes of the American Astronomical Society, 5, 109

\bibitem[{{Mathur}(2000)}]{2000MNRAS.314L..17M}
{Mathur}, S. 2000, \mnras, 314, L17

\bibitem[{{Meenakshi} {et~al.}(2022){Meenakshi}, {Mukherjee}, {Wagner},
  {Nesvadba}, {Bicknell}, {Morganti}, {Janssen}, {Sutherland}, \&
  {Mandal}}]{2022MNRAS.516..766M}
{Meenakshi}, M., {Mukherjee}, D., {Wagner}, A.~Y., {et~al.} 2022, \mnras, 516,
  766

\bibitem[{{Mullaney} {et~al.}(2013){Mullaney}, {Alexander}, {Fine}, {Goulding},
  {Harrison}, \& {Hickox}}]{2013MNRAS.433..622M}
{Mullaney}, J.~R., {Alexander}, D.~M., {Fine}, S., {et~al.} 2013, \mnras, 433,
  622

\bibitem[{{Nagar} {et~al.}(2002){Nagar}, {Oliva}, {Marconi}, \&
  {Maiolino}}]{2002A&A...391L..21N}
{Nagar}, N.~M., {Oliva}, E., {Marconi}, A., \& {Maiolino}, R. 2002, \aap, 391,
  L21

\bibitem[{{Ojha} {et~al.}(2021){Ojha}, {Chand}, \&
  {Gopal-Krishna}}]{2021MNRAS.501.4110O}
{Ojha}, V., {Chand}, H., \& {Gopal-Krishna}. 2021, \mnras, 501, 4110

\bibitem[{{Ojha} {et~al.}(2022){Ojha}, {Jha}, {Chand}, \&
  {Singh}}]{2022MNRAS.514.5607O}
{Ojha}, V., {Jha}, V.~K., {Chand}, H., \& {Singh}, V. 2022, \mnras, 514, 5607

\bibitem[{{Olgu{\'\i}n-Iglesias} {et~al.}(2020){Olgu{\'\i}n-Iglesias},
  {Kotilainen}, \& {Chavushyan}}]{2020MNRAS.492.1450O}
{Olgu{\'\i}n-Iglesias}, A., {Kotilainen}, J., \& {Chavushyan}, V. 2020, \mnras,
  492, 1450

\bibitem[{{Osterbrock} \& {Ferland}(2006)}]{2006agna.book.....O}
{Osterbrock}, D.~E. \& {Ferland}, G.~J. 2006, {Astrophysics of gaseous nebulae
  and active galactic nuclei} (University Science Books)

\bibitem[{{Osterbrock} \& {Koski}(1976)}]{1976MNRAS.176P..61O}
{Osterbrock}, D.~E. \& {Koski}, A.~T. 1976, \mnras, 176, 61P

\bibitem[{{Osterbrock} \& {Pogge}(1985)}]{1985ApJ...297..166O}
{Osterbrock}, D.~E. \& {Pogge}, R.~W. 1985, \apj, 297, 166

\bibitem[{{Panda} {et~al.}(2020){Panda}, {Marziani}, \&
  {Czerny}}]{2020CoSka..50..293P}
{Panda}, S., {Marziani}, P., \& {Czerny}, B. 2020, Contributions of the
  Astronomical Observatory Skalnate Pleso, 50, 293

\bibitem[{{Popovic}(2006)}]{2006SerAJ.173....1P}
{Popovic}, L.~C. 2006, Serbian Astronomical Journal, 173, 1

\bibitem[{{Pradhan} {et~al.}(2006){Pradhan}, {Montenegro}, {Nahar}, \&
  {Eissner}}]{2006MNRAS.366L...6P}
{Pradhan}, A.~K., {Montenegro}, M., {Nahar}, S.~N., \& {Eissner}, W. 2006,
  \mnras, 366, L6

\bibitem[{{Raki{\'c}}(2022)}]{2022MNRAS.516.1624R}
{Raki{\'c}}, N. 2022, \mnras, 516, 1624

\bibitem[{{Rawlings} \& {Saunders}(1991)}]{1991Natur.349..138R}
{Rawlings}, S. \& {Saunders}, R. 1991, \nat, 349, 138

\bibitem[{{Riess} {et~al.}(2022){Riess}, {Yuan}, {Macri}, {Scolnic}, {Brout},
  {Casertano}, {Jones}, {Murakami}, {Anand}, {Breuval}, {Brink}, {Filippenko},
  {Hoffmann}, {Jha}, {D'arcy Kenworthy}, {Mackenty}, {Stahl}, \&
  {Zheng}}]{2022ApJ...934L...7R}
{Riess}, A.~G., {Yuan}, W., {Macri}, L.~M., {et~al.} 2022, \apjl, 934, L7

\bibitem[{{Romano} {et~al.}(2023){Romano}, {L{\"a}hteenm{\"a}ki}, {Vercellone},
  {Foschini}, {Berton}, {Raiteri}, {Braito}, {Ciroi}, {J{\"a}rvel{\"a}},
  {Baitieri}, {Varglund}, {Tornikoski}, \& {Suutarinen}}]{2023A&A...673A..85R}
{Romano}, P., {L{\"a}hteenm{\"a}ki}, A., {Vercellone}, S., {et~al.} 2023, \aap,
  673, A85

\bibitem[{{Schmitt} {et~al.}(2003{\natexlab{a}}){Schmitt}, {Donley},
  {Antonucci}, {Hutchings}, \& {Kinney}}]{2003ApJS..148..327S}
{Schmitt}, H.~R., {Donley}, J.~L., {Antonucci}, R.~R.~J., {Hutchings}, J.~B.,
  \& {Kinney}, A.~L. 2003{\natexlab{a}}, \apjs, 148, 327

\bibitem[{{Schmitt} {et~al.}(2003{\natexlab{b}}){Schmitt}, {Donley},
  {Antonucci}, {Hutchings}, {Kinney}, \& {Pringle}}]{2003ApJ...597..768S}
{Schmitt}, H.~R., {Donley}, J.~L., {Antonucci}, R.~R.~J., {et~al.}
  2003{\natexlab{b}}, \apj, 597, 768

\bibitem[{Schönebeck {et~al.}(2014)Schönebeck, Puzia, Pasquali, Grebel,
  Kissler-Patig, Kuntschner, Lyubenova, \& Perina}]{2014A&A...572A..13S}
Schönebeck, F., Puzia, T.~H., Pasquali, A., {et~al.} 2014, \aap, 572, A13

\bibitem[{{Seyfert}(1943)}]{1943ApJ....97...28S}
{Seyfert}, C.~K. 1943, \apj, 97, 28

\bibitem[{{{\'S}niegowska} {et~al.}(2023){{\'S}niegowska}, {Panda}, {Czerny},
  {Savi{\'c}}, {Mart{\'\i}nez-Aldama}, {Marziani}, {Wang}, {Du}, {Popovi{\'c}},
  \& {Saraf}}]{2023A&A...678A..63S}
{{\'S}niegowska}, M., {Panda}, S., {Czerny}, B., {et~al.} 2023, \aap, 678, A63

\bibitem[{{Urry} \& {Padovani}(1995)}]{1995PASP..107..803U}
{Urry}, C.~M. \& {Padovani}, P. 1995, \pasp, 107, 803

\bibitem[{{Varglund} {et~al.}(2023){Varglund}, {J{\"a}rvel{\"a}}, {Ciroi},
  {Berton}, {Congiu}, {L{\"a}hteenm{\"a}ki}, \& {Di
  Mille}}]{2023A&A...679A..32V}
{Varglund}, I., {J{\"a}rvel{\"a}}, E., {Ciroi}, S., {et~al.} 2023, \aap, 679,
  A32

\bibitem[{{Veilleux} \& {Osterbrock}(1987)}]{1987ApJS...63..295V}
{Veilleux}, S. \& {Osterbrock}, D.~E. 1987, \apjs, 63, 295

\bibitem[{{Weilbacher} {et~al.}(2020){Weilbacher}, {Palsa}, {Streicher},
  {Bacon}, {Urrutia}, {Wisotzki}, {Conseil}, {Husemann}, {Jarno}, {Kelz},
  {P{\'e}contal-Rousset}, {Richard}, {Roth}, {Selman}, \&
  {Vernet}}]{2020A&A...641A..28W}
{Weilbacher}, P.~M., {Palsa}, R., {Streicher}, O., {et~al.} 2020, \aap, 641,
  A28

\bibitem[{{Williams} {et~al.}(2002){Williams}, {Pogge}, \&
  {Mathur}}]{2002AJ....124.3042W}
{Williams}, R.~J., {Pogge}, R.~W., \& {Mathur}, S. 2002, \aj, 124, 3042

\bibitem[{{Xin} {et~al.}(2022){Xin}, {Xiong}, {Bai}, {Liu}, {Lu}, \&
  {Mao}}]{2022RAA....22g5001X}
{Xin}, Y.-X., {Xiong}, D.-R., {Bai}, J.-M., {et~al.} 2022, Research in
  Astronomy and Astrophysics, 22, 075001

\bibitem[{{Yao} \& {Komossa}(2023)}]{2023MNRAS.523..441Y}
{Yao}, S. \& {Komossa}, S. 2023, \mnras, 523, 441

\bibitem[{{Zhang} {et~al.}(2017){Zhang}, {Zhang}, {Zhu}, {Yi}, {Yao}, {Lu}, \&
  {Liang}}]{2017ApJ...849...42Z}
{Zhang}, J., {Zhang}, H.-M., {Zhu}, Y.-K., {et~al.} 2017, \apj, 849, 42

\bibitem[{{Zhou} {et~al.}(2003){Zhou}, {Wang}, {Dong}, {Zhou}, \&
  {Li}}]{2003ApJ...584..147Z}
{Zhou}, H.-Y., {Wang}, T.-G., {Dong}, X.-B., {Zhou}, Y.-Y., \& {Li}, C. 2003,
  \apj, 584, 147

\end{thebibliography}

\begin{appendix} 
\onecolumn
\section{Fitting parameters}
In this section, we present the fitting parameters of the different lines in the SDSS, X-Shooter, and MUSE spectra (see Table~\ref{tab_a1}). 

By comparing the flux of the [O II]$\lambda\lambda$3726,3729 lines, we obtain a ratio of approximately 1.4--1.2. Some authors (\citeads{2006MNRAS.366L...6P};\citeads{2013A&A...559A..18C}) have found a typical range of 1.5 to 0.35 for this ratio, where the limit values correspond to low ($\rightarrow 0$) and high ($\rightarrow \infty$) electron densities, respectively. From our result, we can infer that the [O II] gas is associated with an extremely low electron density. 

Furthermore, due to the X-Shooter spectral coverage, it is possible to obtain {two additional properties of the system from} H$\alpha$ and [N II]$\lambda$6583: the Seyfert type (discussed in Sections~\ref{sec3.3},~\ref{sec4.1}) and the classification from the Baldwin, Phillips, and Terlevich/Veilleux \& Osterbrock (BPT/VO) diagram (\citeads{1981PASP...93....5B};\citeads{1987ApJS...63..295V}). The BPT/VO diagram relates H$\alpha$ and H$\beta$ in their narrow components, to [O III]$\lambda$5007 and [N II]$\lambda$6583. In this case, following\citeads{2003MNRAS.346.1055K};\citeads{2006MNRAS.372..961K} classifications, PMN J0948+0022 is an AGN that lies close to the separation between Seyfert and Low-Ionization Nuclear Emission-line Regions. The obtained ratios are: log([N II]/H$\alpha$)=(-0.38$\pm$0.05) and log([O III]/H$\beta$)=(-0.01$\pm$0.12).

\begin{table}[h!]
\centering
\caption{List of the most important identified lines.}
    \begin{tabular}{llll}
            \hline
        {\bf Line/Component} & {\bf Symbol} & {\bf Flux [$10^{-17}$ erg/(s$\cdot$cm$^2$)]} & {\bf FWHM [km s$^{-1}$]} \\
            \hline
            \hline
                        	 \multicolumn{4}{c}{SDSS} \\
	    \hline
            \hline
		[O II] 							&	[O II]$\lambda$3727   					&	34 $\pm$ 5		& < 370\\
		H$\beta$ narrow component 			&	H$\beta_{\rm n}$   						& 	36 $\pm$ 6	 	& < 370*	\\
		H$\beta$ total broad component 		&	H$\beta_{\rm b}$						&	345 $\pm$ 37		& 1529 $\pm$ 96 \\
		H$\beta$ broad component 1			&	H$\beta_{\rm b1}$						&	154 $\pm$ 27		& 2908 $\pm$ 274 \\
		H$\beta$ broad component 2 			&	H$\beta_{\rm b2}$						&	190 $\pm$ 24		& 1257 $\pm$ 94 \\
		H$\beta$ total (4861\AA)				&	H$\beta_{\rm tot}$						&	380 $\pm$ 37		& 1095 $\pm$ 64 \\
		$\mathrm{[}$O III] core component	 	&	$\mathrm{[}$O III]$\lambda$5007$_{\rm c}$	&	< 7				& < 370*	 	 \\
		$\mathrm{[}$O III] blue wing component	&	$\mathrm{[}$O III]$\lambda$5007$_{\rm o}$	&	< 15				& < 800  	 \\
             \hline
            \hline
                        	 \multicolumn{4}{c}{X-Shooter} \\
	    \hline
            \hline
		$\mathrm{[}$O II] 					&	$\mathrm [$O II]$\lambda$3726   			&  	3.0 $\pm$ 0.6		& 95 $\pm$ 27 \\
		$\mathrm{[}$O II] 					&	$\mathrm [$O II]$\lambda$3729   			&  	4.3 $\pm$ 0.4		& 73 $\pm$ 11 \\
		H$\beta$ narrow component 			&	H$\beta_{\rm n}$   						&  	10.5 $\pm$ 0.4		& 90*  \\
		H$\beta$ total broad component 		&	H$\beta_{\rm b}$						&	362 $\pm$ 16		& 1520 $\pm$ 31 \\
		H$\beta$ broad component 1			&	H$\beta_{\rm b1}$						&	230 $\pm$ 15		& 3586 $\pm$ 73 \\
		H$\beta$ broad component 2			&	H$\beta_{\rm b2}$						&	132 $\pm$ 6		& 1132 $\pm$ 21 \\ 		
		H$\beta$ total (4861\AA)			 	&	H$\beta_{\rm tot}$						&	373 $\pm$ 16		& 866 $\pm$ 35 \\
		$\mathrm{[}$O III] core component	 	&	$\mathrm{[}$O III]$\lambda$5007$_{\rm c}$   	&  	3.2 $\pm$ 0.4		& 90* \\
		$\mathrm{[}$O III] blue wing component	&	$\mathrm{[}$O III]$\lambda$5007$_{\rm o}$	&	17 $\pm$ 2		& 663 $\pm$ 71 \\	
		H$\alpha$ narrow component 			&	H$\alpha_{\rm n}$   						&  	22 $\pm$ 1		& 90*  \\
		H$\alpha$ total broad component 		&	H$\alpha_{\rm b}$						&	960 $\pm$ 36		& 1466 $\pm$ 20 \\
		H$\alpha$ broad component 1			&	H$\alpha_{\rm b1}$						&	518 $\pm$ 33		& 3363 $\pm$ 59 \\
		H$\alpha$ broad component 2			&	H$\alpha_{\rm b2}$						&	442 $\pm$ 14		& 1181 $\pm$ 14 \\ 		
		H$\alpha$ total (6563\AA)			 	&	H$\alpha_{\rm tot}$						&	982 $\pm$ 36		& 1010 $\pm$ 22 \\
		$\mathrm{[}$N II] 				 	&	$\mathrm{[}$N II]$\lambda$6583		   	&  	9.1 $\pm$ 0.8		& 90* \\	
             \hline
            \hline
                        	 \multicolumn{4}{c}{MUSE (all observations)} \\
	    \hline
            \hline
		[O II] 							&	$\mathrm [$O II]$\lambda$3726   			& 	8.4 $\pm$ 0.6		& 158 $\pm$ 18 \\
		$\mathrm{[}$O II] 					&	$\mathrm [$O II]$\lambda$3729   			&  	9.9 $\pm$ 0.6		& 146 $\pm$ 15 \\
		H$\beta$ narrow component 			&	H$\beta_{\rm{n}}$   						&   	9.2 $\pm$ 0.3		& 136* \\
		H$\beta$ total broad component 		&	H$\beta_{\rm{b}}$						& 	353 $\pm$ 7		& 1681 $\pm$ 17 \\
		H$\beta$ broad component 1 			&	H$\beta_{\rm{b1}}$						& 	230 $\pm$ 6		& 3948 $\pm$ 36 \\
		H$\beta$ broad component 2			&	H$\beta_{\rm{b2}}$						& 	122 $\pm$ 3		& 1226 $\pm$ 12  \\
		H$\beta$ total (4861\AA) 			 	&	H$\beta_{\rm{tot}}$						& 	362 $\pm$ 7		& 1239 $\pm$ 17 \\
		$\mathrm{[}$O III] core component	 	&	$\mathrm{[}$O III]$\lambda$5007$_{\rm{c}}$   	&   	6.6 $\pm$ 0.2 		& 136* \\
		$\mathrm{[}$O III] blue wing component	&	$\mathrm{[}$O III]$\lambda$5007$_{\rm{o}}$	& 	12 $\pm$ 1		& 634 $\pm$ 48 \\
            \hline
    \end{tabular}
    \label{tab2}
    \tablefoot{ The columns correspond to: name of the line or of the component, the flux, and the FWHM. The "*" symbol refers to those FWHM associated with [O II]$\lambda$3727. Parameters corrected for the instrumental values.}
\end{table}

\begin{table}[!ht]
\caption{Fitting parameters for the main spectral lines and the observed redshift (z$_{\rm obs}$) for those lines identified at least at 5$\sigma$.}
\centering
	\begin{tabular}{llllll}
            \hline
		{\bf Component} & {\bf Amplitude} & {\bf Center} & {\bf Width ($\sigma_{\rm gau}$)} & {\bf z$_{\rm obs}$} \\
		 & {\bf {\scriptsize{[$10^{-17}$ erg/(s$\cdot$cm$^{2}\cdot\AA$)]}}} & {\bf [$\AA$]} & {\bf [$\AA$] } & \\
            \hline
            \hline
            	 \multicolumn{5}{c}{SDSS} \\
	    \hline
            \hline
            	H$\beta_{\rm n}$						& -			     & $\sim$ 4340.69 	& - 					& 0.58573\\
		$\mathrm{[}$O II]$\lambda$3727			& 5.63 $\pm$ 0.71 & 3727.39 $\pm$ 0.06 & 0.56 $\pm$ 0.46 		& 0.58557\\  
            	H$\beta_{\rm n}$						& 4.69 $\pm$ 0.91 & 4861.78 $\pm$ 0.40 & 1.13* 				& 0.58571\\
		H$\beta_{{\rm{b1}}}$						& 3.23 $\pm$ 0.54 & 4859.20 $\pm$ 1.75 & 20.01 $\pm$ 0.04 		& -\\
		H$\beta_{{\rm{b2}}}$						& 9.18 $\pm$ 0.82 & 4861.11 $\pm$ 0.56 & 8.09 $\pm$ 0.81 		& -\\
		$\mathrm{[}$O III]$\lambda$5007$_{{\rm c}}$	& 1.54 $\pm$ 0.54 & 5007.15 $\pm$ 0.33 & 0.75* 				& -\\
		$\mathrm{[}$O III]$\lambda$5007$_{{\rm o}}$	& 0.83 $\pm$ 0.31 & 5000.18 $\pm$ 0.63 & 4.50$^@$ 			& -\\
	    \hline
            \hline
            	 \multicolumn{5}{c}{X-Shooter} \\
	    \hline
            \hline
            	$\mathrm{[}$O II]$\lambda$3726			& 1.86 $\pm$ 0.30 & 3726.26 $\pm$ 0.16 & 0.50 $\pm$ 0.10 	& 0.58502\\  
            	$\mathrm{[}$O II]$\lambda$3729			& 3.14 $\pm$ 0.29 & 3728.78 $\pm$ 0.06 & 0.39 $\pm$ 0.01 	& 0.58508\\
		H$\delta$								&  -			     & $\sim$  4101.32 	& - 				& 0.58488\\
            	H$\beta_{{\rm n}}$						& 5.47 $\pm$ 0.31 & 4861.27 $\pm$ 0.04 & 0.58* 			& 0.58508 \\
		H$\beta_{{\rm{b1}}}$						& 3.71 $\pm$ 0.21 & 4862.33 $\pm$ 0.50 & 24.78 $\pm$ 0.78 	& - \\
		H$\beta_{{\rm{b2}}}$						& 6.69$\pm$ 0.20  & 4860.59 $\pm$ 0.15 & 7.87 $\pm$ 0.28 	& - \\
		$\mathrm{[}$O III]$\lambda$5007$_{{\rm c}}$	& 1.61 $\pm$ 0.27 & 5006.61 $\pm$ 0.20 & 0.60* 			& 0.58496 \\
		$\mathrm{[}$O III]$\lambda$5007$_{{\rm o}}$	& 1.40 $\pm$ 0.10 & 5000.36 $\pm$ 0.51 & 4.70 $\pm$ 0.35 	& - \\
            	H$\alpha_{{\rm n}}$						& 8.97 $\pm$ 0.45 & 6562.47 $\pm$ 0.07 & 0.78* 			& 0.58502\\
		H$\alpha_{{\rm{b1}}}$					& 6.50 $\pm$ 0.37 & 6563.47 $\pm$ 0.55 & 31.49 $\pm$ 0.90 	& - \\
		H$\alpha_{{\rm{b2}}}$					& 16.02 $\pm$ 0.35 & 6561.91 $\pm$ 0.13 & 11.07 $\pm$ 0.25 & - \\
		$\mathrm{[}$N II]$\lambda$6583			& 3.71 $\pm$ 0.39 & 6583.38 $\pm$ 0.12 & 0.78* 			& 0.58512\\
		$\mathrm{[}$S II]$\lambda$6716			&  -			     & $\sim$  6717.18 	& - 				& 0.58521\\
            \hline
            \hline
            	 \multicolumn{5}{c}{MUSE} \\
	    \hline
            \hline
		$\mathrm{[}$O II]$\lambda$3726			& 3.20 $\pm$ 0.14 & 3726.17 $\pm$ 0.07 	& 0.83 $\pm$ 0.10 	& 0.58536 \\    
		$\mathrm{[}$O II]$\lambda$3729			& 3.96 $\pm$ 0.15 & 3728.92 $\pm$ 0.07 	& 0.77 $\pm$ 0.08 	& 0.58511 \\            
            	H$\epsilon$							&  -			     & $\sim$  3970.35 	& - 				& 0.58508\\
            	H$\delta$								&  -			     & $\sim$  4102.15 	& - 				& 0.58513\\
            	H$\gamma$							&  -			     & $\sim$  4341.27 	& - 				& 0.58526\\
		H$\beta_{{\rm n}}$						& 2.96 $\pm$ 0.13 & 4861.72 $\pm$ 0.04 & 0.93* 			& 0.58519 \\
		H$\beta_{{\rm{b1}}}$						& 3.38 $\pm$ 0.08 & 4860.96 $\pm$ 0.25 & 27.19 $\pm$ 0.35 	& - \\
		H$\beta_{{\rm{b2}}}$						& 5.75 $\pm$ 0.08 & 4861.64 $\pm$ 0.08 & 8.44 $\pm$ 0.14 	& - \\
		$\mathrm{[}$O III]$\lambda$5007$_{{\rm c}}$	& 2.04 $\pm$ 0.09 & 5006.86 $\pm$ 0.06 & 0.96* 			& 0.58513\\
		$\mathrm{[}$O III]$\lambda$5007$_{{\rm o}}$	& 1.04 $\pm$ 0.06 & 4997.11 $\pm$ 0.34 & 4.45 $\pm$ 0.28 	& - \\
		$\mathrm{[}$N I]$\lambda$5200			&  -			     & $\sim$  5198.07 	& - 				& 0.58430\\
            \hline
	\end{tabular}
	\label{tab_a1}
	\tablefoot{The widths highlighted by an * do not indicate an error because they are fixed in the fitting procedure, see Section~\ref{sec3.2} for the details. The [O III] outflow width in the SDSS case, highlighted with $^@$, is fixed to the average value of the other two cases.}
\end{table}

\FloatBarrier

\section{Telluric contamination in H$\beta$}
\begin{figure}[!ht]
    \centering
    \includegraphics[width=0.55\textwidth]{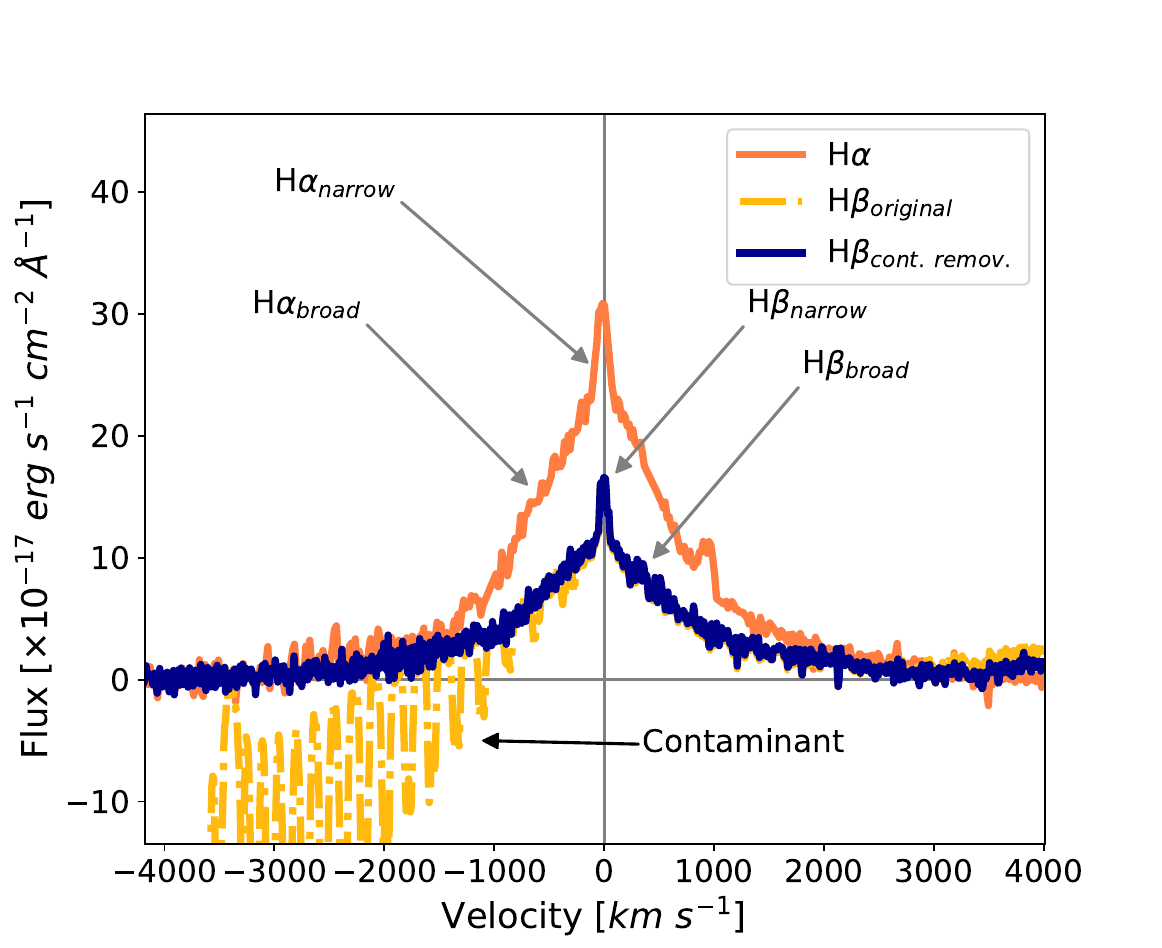} 
    \caption{Comparison between H$\beta$ (in blue for the corrected profile, and in dotted-yellow for the contaminated one) and H$\alpha$ (in orange) in the X-Shooter spectrum. The grey arrows refer to the broad and narrow components of H$\beta$ and H$\alpha$. The black arrow stresses the position of the telluric absorption.}
    \label{fig_a1}
\end{figure}

The original shape of H$\beta$ is affected by significant telluric absorption in the blue part (see Fig.~\ref{fig_a1}). To correct for this contribution, we assumed a symmetric shape for H$\beta$, based on that of H$\alpha$. Consequently, we mirrored the right side of the line onto the left side and added random noise proportional to what was observed in the 5050--5150~\AA~range. The entire procedure was performed using \texttt{IRAF} and specific Python codes. 

\FloatBarrier

\section{Other sources in the MUSE FoV}
\begin{figure}[ht!]
    \centering
    \includegraphics[width=0.45\textwidth]{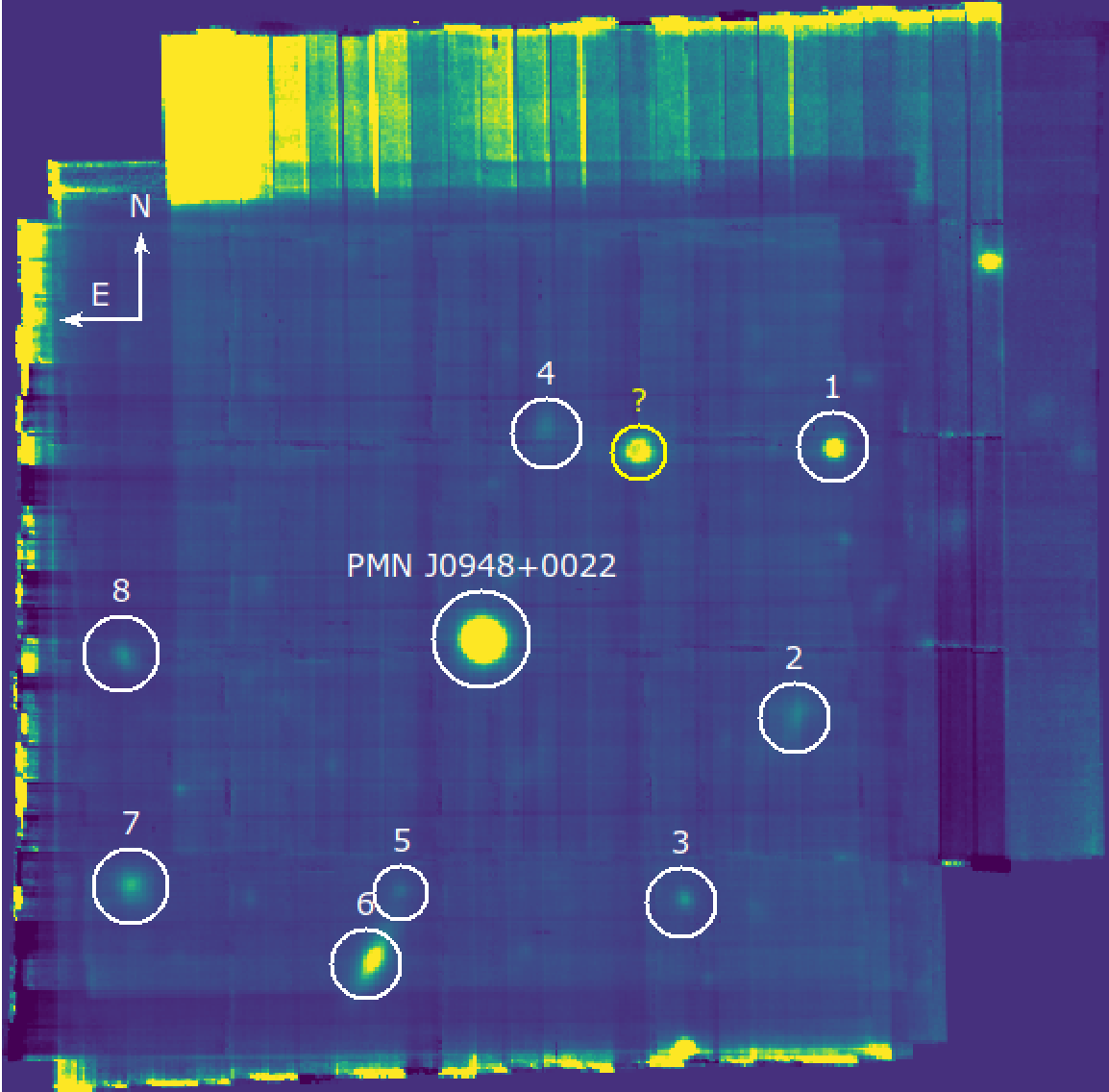} 
    \caption{MUSE WFM-NOAO FoV (1x1 arcmin$^2$) of PMN J0948+0022 (at the center). The MUSE scale is of 0.2 arcsec/pixel, combined with the scale of PMN J0948+0022 of s=6.31 kpc/arcsec, we find $\sim$ 1.26 kpc/pixel for the central source. We averaged the z-axis (spectral axis) to enhance the flux for each pixel of the image. The white circles we highlight the sources for which we measured the spectroscopic redshift. The numbers correspond to those listed in the first column of Table~\ref{tab_a2}.}
    \label{fig_a2}
\end{figure}

As an IFS, MUSE produces data cubes that can be separated in two components: one for the two spatial directions (x,y) (see for example the FoV of PMN J0948+0022 in Fig.~\ref{fig_a2}), and one spectral direction (z) for each pixel. In Table~\ref{tab_a2} we present the first measurement of the spectroscopic redshift of the sources in the MUSE FoV of PMN J0948+0022. The selection of the sources was made applying a filter to the original data cube, limiting those sources with a flux smaller than $4\times10^{-20}$ erg s$^{-1}$ cm$^{-2}$. This threshold corresponds to $\sim2$ times the noise level, calculated in an empty region of the image (top-left part of Fig.~\ref{fig_a2}). All the objects that we present are galaxies, except for [VV2006] J094855.7+002238. It is also important to stress the presence of a possible artifact, highlighted by the yellow circle in Fig.~\ref{fig_a2}. In this position none of the public catalogs present a source, so we classified it as an instrumental effect. The resulting redshift is the weighted average of the redshifts for all the identified lines in the corresponding spectra.

\begin{sidewaystable}
\caption{Objects in the MUSE FoV.}
	\begin{tabular}{lllllll}
            \hline
		 & {\bf Name} & {\bf RA (J2000)} & {\bf DEC (J2000)} & {\bf Public redshift} & {\bf New redshift} & {\bf Identified lines} \\
		  & & {\bf [h m s]} & {\bf [$^{\circ}$ ' '']} & & & \\
	    \hline
	    \hline
	    1 & [VV2006] J094855.7+002238				      & 09 48 55.613 & +00 22 39.214 & z$_{\rm s}$=1.6$^a$ & (1.5995$\pm$0.0036) &  Si III]$\lambda$1892, Mg II]$\lambda$2799 \\	    
	    & & & & & &  \\
	    \hline
	    2 & WiggleZ S09J094855793+00222011		      & 09 48 55.795 & +00 22 20.092 & z$_{\rm s}$=0.07$^b$ & (0.0696$\pm$0.0001) & H$\beta$, [O III]$\lambda\lambda$4959,5007, \\
	    & & & & & &  H$\alpha$, [N II]$\lambda\lambda$6548,6583,\\
	     & & & & & &  [S II]$\lambda\lambda$6717,6731\\
	    \hline
	    3 & SDSS J094856.34+002206.6				      & 09 48 56.348 & +00 22 06.699 & z$_{\rm p}$=0.83$^c$ & (0.4922$\pm$0.0001) & [O III]$\lambda$5007, [N II]$\lambda$6548, \\
	    & & & & & &  and a very faint H$\beta$ \\
	    \hline
	    4 & SDSS J094857.03+002240.3				      & 09 48 57.034 & +00 22 40.389 & z$_{\rm p}$=0.46$^d$  & (0.4922$\pm$0.0010) & [O III]$\lambda\lambda$4959,5007, \\
	    & & & & & & a faint H$\beta$ \\
	    \hline
	    5 & ID 8000332396004246 * 					      & 09 48 57.721 & +00 22 07.598 & z$_{\rm p}$=0.96$^c$ & (0.7600$\pm$0.0001) & [O II]$\lambda$3727, \\
	    & & & & & & [O III]$\lambda\lambda$4859,5007, H$\beta$ \\
	    \hline
	    6 & SDSS J094857.85+002202.2				      & 09 48 57.859 & +00 22 02.245 & z$_{\rm p}$=0.30$^e$ & (0.2830$\pm$0.0001) & [O II]$\lambda$3727, H$\beta$, \\
	    & & & & & & [O III]$\lambda\lambda$4959,5007, H$\alpha$, \\
	    & & & & & & [N II]$\lambda\lambda$6548,6583, \\
	    & & & & & & [S II]$\lambda\lambda$6717,6731\\
	    \hline
	   7 & SDSS J094859.00+002207.9				      & 09 48 59.009 & +00 22 07.954 & z$_{\rm p}$=0.25$^e$ & (0.1821$\pm$0.0001) & H$\beta$, [O III]$\lambda\lambda$4959,5007, \\
	   & & & & & & H$\alpha$, [N II]$\lambda\lambda$6548,6583, \\
	    & & & & & & [S II]$\lambda\lambda$6717,6731\\
	    \hline
	   8 & SDSS J094859.03+002224.3				      & 09 48 59.034 & +00 22 24.334 & z$_{\rm p}$=0.77$^c$,0.48$^e$ & (0.3732$\pm$0.0001) & [O II]$\lambda$3727, H$\beta$, \\
	   & & & & & & H$\alpha$, [N II]$\lambda\lambda$6548,6583, \\
	   & & & & & & [S II]$\lambda\lambda$6717,6731 \\
	 \hline
	\end{tabular}
	\label{tab_a2}
	\tablefoot{The columns correspond to: the identification number used in Fig.~\ref{fig_a2}, the name of the source (the one with * is from DESI-DR8 VII/292/south), coordinates of the object (RA, DEC), the public redshift (spectroscopic, z$_{\rm s}$, or photometric, z$_{\rm p}$) with the corresponding reference, the new redshift from the MUSE data cube, and the identified lines to calculate the redshift. References: a)\citeads{2009ApJS..182..543A}, b)\citeads{2018MNRAS.474.4151D}, c)\citeads{2022MNRAS.512.3662D}, d)\citeads{2020ApJS..249....3A}, e)\citeads{2015ApJS..219...12A}.}
\end{sidewaystable}

\end{appendix}

\end{document}